\documentclass[12pt,lettersize,journal,onecolumn,draftclsnofoot]{IEEEtran}

\usepackage[numbers,sort&compress]{natbib}

\usepackage{amsmath, amssymb}
\usepackage{algorithm}
\usepackage{algorithmic}
\usepackage{bbm}
\usepackage{array}
\usepackage[usenames,dvipsnames]{color}
\usepackage{url}
\usepackage{hyperref}
\usepackage{cuted}
\usepackage{graphicx}
\usepackage{afterpage}
\usepackage{epstopdf}
\usepackage[font=footnotesize,labelfont=bf]{caption}
\usepackage{subcaption}
\usepackage{soul}

\begin{document}

\title{Minimizing Energy Consumption based on Waste Factor in Multi-user Multi-relay Systems}

\author{Nurullah~Sevim,\thanks{This material is based upon work supported in part by the U.S. Department of Energy, Office of Science, Office of Advanced Scientific Computing Research, Early Career Research Program under Award Number DE-SC0023957, in part by the National Science Foundation under Grant No. 2535873, and in part by the Hagler Institute for Advanced Study at Texas A\&M University.}\thanks{(\textit{*Corresponding author: Nurullah~Sevim}.)}
\thanks{Nurullah~Sevim is with the Department of Electrical \& Computer Engineering, Texas A\&M University, Texas, USA (E-mail: nurullahsevim@tamu.edu).}Mostafa~Ibrahim\thanks{Mostafa~Ibrahim is with the Department of Engineering Technology and Industrial Distribution, Texas A\&M University, Texas, USA (E-mail: mostafa.ibrahim@tamu.edu).},
Sabit~Ekin,~\IEEEmembership{Senior~Member,~IEEE}\thanks{Sabit~Ekin  is with the Departments of Engineering Technology and Industrial Distribution, and Electrical \& Computer Engineering, Texas A\&M University, Texas, USA (E-mail: sabitekin@tamu.edu).}, and~Theodore~S.~Rappaport,~\IEEEmembership{Fellow,~IEEE}\thanks{Theodore~S.~Rappaport  is with  New York University (NYU), and the founding director of NYU WIRELESS, Tandon School of Engineering, Brooklyn, NY, USA (E-mail: tr51@nyu.edu).}}

\maketitle

\begin{abstract}

The escalating energy demands of modern wireless networks make energy-efficient system design a central challenge for future 6G deployments. Relay-assisted communication has long been recognized as a powerful tool for improving coverage and throughput, tracing its roots to early repeater communication systems and the foundational relay channel models in information theory~\cite{cover79relay}. Meanwhile, modern relay selection and routing strategies are primarily driven by rate-based or transmit-power-centric metrics, which often neglect hardware-level losses that is independent of the transmitted signal. As networks grow denser and more heterogeneous, signal-independent hardware inefficiencies and circuit power consumption increasingly dominate the total energy consumption and fundamentally alter the optimality of relay assignment decisions.

To address the challenge of modeling signal-independent power consumption in relay-assisted systems, the Waste Factor (WF) framework is adopted: a hardware-aware, unifying metric that explicitly captures signal-independent power losses across cascaded transceivers and wireless channels. Using WF, the energy consumption of multi-user, multi-relay wireless systems is quantified and minimized under practical hardware and network constraints.
The optimal relay–user assignment problem is formulated as an energy-minimization problem subject to communication feasibility and quality-of-service constraints, ensuring that each user achieves a minimum required received power (equivalently, a target link reliability). The formulation jointly accounts for component inefficiencies, propagation losses, and relay-selection constraints. 
The proposed framework demonstrates that the WF formulation provides a fundamentally new way to model relay-assisted communication, enabling hardware-aware energy-per-bit minimization and a structured representation that enables computationally efficient analysis. Simulation results confirm that WF-based relay assignment reduces total energy per bit while improving the ability of users to satisfy minimum received-power requirements, outperforming conventional formulations across both metrics.
\end{abstract}


\section{Introduction}

The global demand for digital communication continues to grow at an unprecedented pace as wireless networks evolve to support increasingly dense deployments, higher data rates, and a vast ecosystem of connected devices~\cite{iea2025energyai}. Emerging sixth-generation (6G) communication services, including immersive extended reality, autonomous systems, and massive machine-type communication, will require orders-of-magnitude improvements in both throughput and coverage. However, this rapid technological expansion is accompanied by escalating energy consumption across information and communication technologies (ICT)~\cite{iea_data_centres}. Recent estimates suggest that ICT already accounts for a measurable share of global electricity consumption, a trend expected to intensify as networks become more heterogeneous and computation-intensive~\cite{9810097}. Improving the energy efficiency of communication systems has therefore become a central research priority, both to reduce operational expenditures and to ensure the environmental sustainability of future wireless infrastructures.

While conventional wireless design emphasizes spectral efficiency and coverage reliability, modern base stations, access points, relays, and user equipment incur substantial signal-independent power consumption from local oscillators, data converters, digital processing, and control circuitry, while power amplifiers and RF front-ends add losses that can dominate total power in low-SNR or low-rate regimes. Classical link-level metrics such as energy per bit or transmit power alone fail to capture this interplay between hardware inefficiency, non-ideal signal chains, and multi-hop routing. Widely used efficiency metrics such as Power Usage Effectiveness (PUE) share this blind spot, as they do not relate energy to useful signal output and can mislead system-level assessments~\cite{rappaport2024wastemj}.

To address the limitations of rate-based and transmit-power-centric metrics in capturing hardware inefficiencies, a recently introduced metric known as the Waste Factor (WF) {-also called Waste Figure when in dB units-} provides a unified framework for modeling power dissipation in cascaded wireless systems~\cite{ying2023waste,rappaport24waste,sevim2025theoretical}. WF aggregates the multiplicative and additive losses incurred across transmitters, receivers, channels, and relays, yielding a hardware-aware representation of the actual energy cost associated with reliable communication. By distinguishing between useful radiated energy {(e.g. energy consumption along the signal path between source and sink)} and signal-independent consumption {(e.g., energy consumption off of the signal path)}, WF enables thorough system-level analyses{ for energy used both in information transmission as well as total  energy consumption in the communication system.}

Relay-assisted wireless communication has been widely investigated as an effective mechanism to improve coverage probability, link reliability, and energy consumption per bit in networks where direct transmission is limited by propagation loss or power constraints. A growing body of work focuses on optimizing relay operation through joint resource allocation and system-level optimization. For instance, \cite{jiang19uav} studies unmanned aerial vehicle (UAV)–relayed wireless systems and formulates joint power and trajectory optimization problems to either maximize the minimum achievable rate or minimize total power consumption under rate constraints. {While such formulations enable flexible system design, they are primarily based on rate maximization or transmit-power minimization and often result in highly coupled non-convex optimization problems. In addition, signal-independent hardware inefficiencies are not explicitly accounted for, thereby limiting the accuracy of energy-consumption modeling. In contrast, a WF-based formulation provides a unified, hardware-aware characterization of energy consumption, offering a more structured representation of relay-assisted communication and potentially reducing modeling complexity by directly operating on end-to-end energy metrics.}

Another line of research focuses on {selecting which relay assists each source–destination transmission and allocating transmit power across the corresponding links,} with the goal of reducing computational complexity and signaling overhead. In \cite{cai08afrelay}, a semi-distributed user relaying algorithm is proposed for amplify-and-forward relay networks, where each relay independently evaluates {whether it can support a given source–destination transmission by comparing its local channel conditions against a predefined threshold. A central controller then selects one relay from the set of feasible candidates to forward the signal.} This approach emphasizes practicality and low complexity by avoiding full global channel state information exchange, while still achieving {achievable-rate performance} close to that of fully centralized relay selection schemes. In contrast, coordinated relay transmission strategies for multi-user cellular systems are investigated in \cite{thai11coordianted}, where joint transmission and interference cancellation principles are used to serve direct and relayed users simultaneously. Optimal scheduling is shown to require exhaustive search, motivating suboptimal yet efficient coordination algorithms that achieve near-optimal system throughput. While these studies demonstrate the benefits of coordination and relay selection, they primarily optimize rate or capacity metrics and do not explicitly model hardware inefficiencies or signal-independent power consumption.

Cooperative transmission from multiple geographically separated transmitters has also been extensively studied in the context of Coordinated Multi-Point (CoMP) communication. In CoMP systems, multiple base stations or transmission points jointly serve a user in order to improve coverage, reduce outage probability, improve received signal power, and boost cell-edge performance. Such coordination can transform otherwise interfering signals into useful contributions at the receiver and significantly improve system throughput and coverage in dense deployments \cite{irmer2011comp,gesbert2010comp}. 

This paper investigates the design of energy-efficient multi-user and multi-relay wireless systems under the WF framework. Conventional relay-assignment approaches are typically based on rate maximization, transmit-power minimization, or shortest-path metrics, which implicitly assume idealized transceivers or treat hardware-related power consumption as a secondary effect. Such formulations often fail to capture the dominant signal-independent power losses arising from practical RF front-ends, baseband processing, and cascaded transceiver chains. These losses become particularly significant in short-range relaying and dense multi-hop deployments. As a result, relay assignments that appear optimal under rate- or power-based metrics can be energy-inefficient when realistic hardware constraints are taken into account.

Unlike prior WF studies~\cite{murdock2013consumption,sevim2025theoretical} that identify analytical conditions for favorable relaying on a single link, this work extends the framework to the network level, jointly addressing relay selection and user assignment. This shift from link-level to network-level minimization yields a fundamentally different problem structure that couples multi-hop cascades, heterogeneous hardware, relay capacity limits, and spatial user distributions, tradeoffs invisible to conventional rate- or shortest-path formulations.

Operating frequency also shapes feasibility. This work considers millimeter-wave (mmWave) communication, a cornerstone of 5G and 6G for its large bandwidths and high data rates. mmWave links suffer severe path loss, limited diffraction, and strong blockage sensitivity, forcing dense deployments, short-range links, and relay assistance to maintain coverage. These same characteristics make mmWave especially sensitive to hardware inefficiency and signal-independent power, amplifying the value of hardware-aware metrics such as WF.

To focus on the structural implications of hardware inefficiencies and relay assignment under the Waste Factor framework, the channel model in this work is treated deterministically through distance-dependent large-scale path loss. This choice isolates first-order propagation effects and enables clear interpretation of the resulting energy trends without the additional variability introduced by stochastic channel fluctuations. Extensions to probabilistic and site-specific propagation models, including log-normal shadow fading, are natural and are left for future work.

Although the WF framework allows several structured special configurations to be expressed in closed form, the general relay-assignment problem exhibits inherent combinatorial complexity. In particular, when the number of users and relays grows, the optimal assignment becomes computationally intractable due to the exponential growth of possible configurations. To overcome the combinatorial complexity and computational intractability of the relay-assignment problem, a suite of methods is employed with complementary strengths: greedy strategies for decoupled regimes, min-cost flow (MCF) reformulation for exact polynomial-time optimization under single-link capacitated assignment, and constructive/Mixed Integer Linear Programming (MILP)-based methods for scalable optimization in multi-relay cooperative regimes. Exhaustive search is included as a benchmark when the network dimensions permit, enabling explicit quantification of the optimality--complexity tradeoff.

The relay-assisted architecture considered in this work bears conceptual similarities to joint-transmission CoMP, where multiple transmission points cooperatively deliver data to a user. However, while the CoMP literature primarily focuses on spectral efficiency and interference mitigation~\cite{li2019comp,zhao2022comp}, this work addresses a complementary problem: the energy-efficient selection of cooperating transmission points under a hardware-aware energy model based on the Waste Factor framework. This energy-centric perspective, based on the WF framework, is particularly relevant for future dense mmWave deployments, where energy consumption and hardware inefficiencies play a critical role.

The major contributions of this work are summarized as follows:

\begin{itemize}
    \item \textbf{WF-based system modeling:} The WF framework is extended to multi-user, multi-relay networks, with explicit end-to-end energy-per-bit expressions for candidate relay paths capturing component inefficiencies, propagation losses, and signal-independent power.

    \item \textbf{Energy-minimizing relay assignment:} The relay-selection and user-assignment problem is formulated as total energy-per-bit minimization under communication feasibility constraints, including minimum received power requirements and relay capacity limits.

    \item \textbf{Exact solutions for structured cases:} Several structured special cases are shown to admit exact solutions, enabling optimal relay assignment in polynomial time.

    \item \textbf{Scalable algorithms with quantified tradeoffs:} For general large-scale systems, scalable sub-optimal algorithms are developed that achieve near-optimal energy-per-bit performance at low computational complexity.

    \item \textbf{Comprehensive evaluation:} Simulations quantify total energy per bit and the fraction of users failing the minimum received power requirement, showing that WF-based assignment reduces energy consumption while improving coverage feasibility over conventional approaches.
\end{itemize}

The remainder of this paper is structured as follows. Section~\ref{sec:preliminaries} reviews the WF model and develops the WF-based energy formulation for multi-hop links. Section~\ref{sec:system_model} introduces the multi-user multi-relay system model, and Section~\ref{sec:problem-form} formulates the relay-assignment energy-minimization problem. Section~\ref{sec:methodology} develops scalable sub-optimal algorithms for the general case. Section~\ref{sec:results} provides simulation results and performance evaluations. Finally, Section~\ref{sec:Conclusion} concludes the paper and outlines directions for future research.

\section{Preliminaries}
\label{sec:preliminaries}

\subsection{Fundamentals of the WF}
\label{sec:waste_factor}

Inspired by the Noise Factor originally introduced by Friis \cite{friis1944noise}, the WF framework provides a systematic approach for quantifying energy inefficiencies in cascaded communication systems. While the Noise Factor characterizes the accumulation of noise across system components, the WF captures how power losses and overhead consumption compound across transmitters, channels, and receivers \cite{ying2023waste,rappaport24waste}.

\begin{figure}
    \centering
    \includegraphics[width=.5\linewidth]{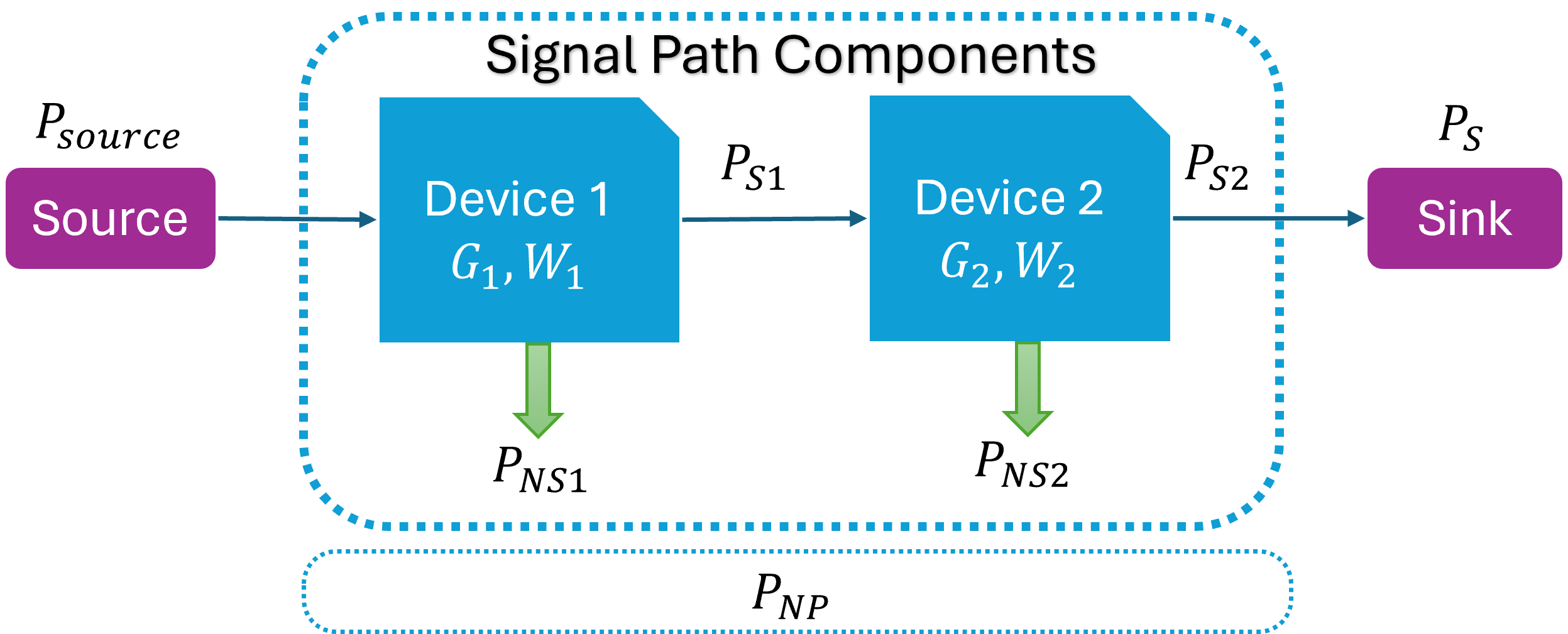}
    \caption{A two-device cascaded communication system depicting the propagation of signal power and the associated power expenditure across successive stages.}
    \label{fig:waste-factor}
\end{figure}

Consider a generic cascaded communication subsystem illustrated in Fig.~\ref{fig:waste-factor} delivering a useful signal power \( P_{\mathrm{S}} \) at its output. The total power consumption of the subsystem can be decomposed as
\begin{equation}
P_{\mathrm{consumed}} = P_{\mathrm{S}} + P_{\mathrm{NS}} + P_{\mathrm{NP}},
\label{eq:power_decomposition}
\end{equation}
where \( P_{\mathrm{NS}} \) denotes signal-supporting power required for modulation, synchronization, and baseband processing, and \( P_{\mathrm{NP}} \) represents non-path-related power consumed by auxiliary circuitry and control logic.

The WF of the subsystem is defined as
\begin{equation}
W \triangleq \frac{P_{\mathrm{S}} + P_{\mathrm{NS}}}{P_{\mathrm{S}}} \geq 1,
\label{eq:waste_factor_def}
\end{equation}
which quantifies the multiplicative overhead incurred per unit of useful received signal power. An ideal, lossless device corresponds to \( W = 1 \).

A cascade of communication components is considered, each characterized by a WF and a gain. Let the cascade consist of \( N \) stages, indexed from the source (\( i=1 \)) to the final receiver (\( i=N \)). Denote by \( W_i \) and \( G_i \) the WF and gain of stage \( i \), respectively. The output signal power of the cascade is denoted by \( P_{\mathrm{S}} \).

Following the WF formalism, the total path-related power consumption of the cascade can be expressed as
\begin{equation}
P_{\mathrm{path}} = W P_{\mathrm{S}},
\label{eq:path_power}
\end{equation}
where \( W \) is the equivalent WF of the cascaded system.

By accounting for the contribution of each stage and normalizing all power terms to the final received signal power, the overall WF of the cascade is given by
\begin{equation}
W
=
1
+
\sum_{k=1}^{N}
\frac{W_k - 1}{\prod_{i=k+1}^{N} G_i}.
\label{eq:waste_factor_cascade}
\end{equation}

In the system model of Section~III, wireless links are characterized by an effective channel gain $G_{ch}$, incorporating large-scale path loss. From an energy-efficiency perspective, the wireless channel is a passive attenuating element and can be modeled as follows:

\begin{equation}
W_{\mathrm{ch}} = \frac{1}{G_{ch}}.
\label{eq:channel_wf}
\end{equation}

Consider a single-hop wireless link consisting of a transmitter, a wireless channel, and a receiver. Let \( W_{\mathrm{tx}} \) and \( W_{\mathrm{rx}} \) denote the WFs of the transmitter and receiver hardware, respectively, and let \( G_{\mathrm{tx}} \) and \( G_{\mathrm{rx}} \) denote their corresponding antenna and RF gains, respectively. Using \eqref{eq:waste_factor_cascade} and \eqref{eq:channel_wf}, the end-to-end WF of the link is
\begin{equation}
W_{\mathrm{link}}
=
W_{\mathrm{rx}}
+
\frac{1}{G_{\mathrm{rx}}}
\left(
\frac{1}{G_{ch}} - 1
\right)
+
\frac{W_{\mathrm{tx}} - 1}{G_{ch}~G_{\mathrm{rx}} } 
=
W_{\mathrm{rx}}
+
\frac{W_{\mathrm{tx}}}{G_{ch}~G_{\mathrm{rx}} }
-
\frac{1}{G_{\mathrm{rx}}}.
\label{eq:link_waste_factor}
\end{equation}

In practical wireless systems, where \( G_{\mathrm{rx}} G_{ch} \ll 1 \), the above expression is dominated by the transmitter and channel terms and can be accurately approximated as
\begin{equation}
W_{\mathrm{link}} \approx \frac{W_{\mathrm{tx}}}{G_{\mathrm{rx}} G_{ch}}.
\label{eq:link_wf_approx}
\end{equation}

Severe channel attenuation causes $W_{\mathrm{link}}$ to grow rapidly, motivating relay-assisted transmission.

\subsection{Energy Consumption and WF}

To quantify energy efficiency, the \emph{Consumption Factor} (CF)~\cite{murdock2013consumption, sevim2025theoretical} is utilized, which measures the maximum achievable data rate per unit of consumed power, mathematically shown as:

\begin{equation}
    CF = \left(\frac{R}{P_{\text{consumed}}}\right)_{max} = \frac{R_{max}}{P_{\text{consumed},min}}.
    \label{cf}
\end{equation}

For an additive white Gaussian noise (AWGN) channel, the maximum achievable data rate is governed by Shannon’s capacity, while the minimum required SNR follows directly from information-theoretic limits. In practical communication systems, however, the total consumed power includes a fixed non-path (circuit) component and a signal-path component that is amplified by hardware inefficiencies. These inefficiencies are captured through the WF, which aggregates power losses across cascaded transceiver chains and wireless links.

In the wideband regime, where the system bandwidth approaches infinity, the minimum consumed energy per bit, denoted by $E_{bc}$, becomes the inverse of the CF. Under this asymptotic condition, the minimum energy per bit consumed by the system admits the closed-form expression
\begin{equation}
    E_{bc}
    = \frac{P_{NP}}{C} + \ln(2)\, N_0 \, W,
    \label{eq:Ebc}
\end{equation}
where $P_{NP}$ denotes the signal-independent (non-path) power consumption, $C$ is the achievable data rate, $N_0$ is the noise power spectral density, and $W$ is the WF of the cascaded transmission path~\cite{murdock2013consumption,sevim2025theoretical}.

\section{System Model}
\label{sec:system_model}

\begin{table}[t]
\centering
\caption{Main Symbols}
\renewcommand{\arraystretch}{1.15}
\setlength{\tabcolsep}{3pt}
\begin{tabular}{ll}
\hline
\textbf{Symbol} & \textbf{Meaning} \\ \hline

$G_{\mathrm{BS},r}$ & Channel gain of BS$\rightarrow$r link \\
$G_{r,k}$ & Channel gain of relay$r\rightarrow$UE$k$ link \\
$G_{\mathrm{BS},k}$ & Channel gain of BS$\rightarrow$UE$k$ link \\

$G_{rx}^\cdot$ & The receiver antenna gain\\
$G_{tx}^\cdot$ & The transmit antenna gain\\

$P_{\mathrm{tx}}^{\mathrm{BS}}$ & BS transmit power \\
$P_{\mathrm{tx}}^{r}$ & Relay transmit power \\
$P_k^{\mathrm{rx}}$ & Received power at user $k$ \\
$P_{\min}$ & Minimum received power requirement \\

$P_{\mathrm{NP}}$ & Signal-independent (non-path) power consumption \\

$W_{\mathrm{BS}},W_r$ & WF of BS and relay transmit chains \\
$W_{\mathrm{UE},k}$ & WF of UE receive chain \\

$E_{bc}$ & Consumed energy per bit \\
$E_{bc}^k(r)$ & Energy per bit serving user $k$ via node $r$ \\
$E_{k,direct}$ & Energy per bit serving user $k$ via direct BS link \\
$E_{bc}^k(\mathcal{S}_k)$ & Energy per bit with serving set $\mathcal{S}_k$ \\
$E_{bc}^k$ & Total consumed energy per bit for user $k$\\

\hline
\end{tabular}
\label{tab:symbols}
\end{table}

A downlink cooperative wireless network is considered, composed of a single base station (BS), a set of $R$ heterogeneous decode-and-forward (DF) relays, and $K$ stationary user equipments (UEs). Each UE is allocated an orthogonal bandwidth $B$, eliminating inter-user interference; this FDMA-like operation isolates energy-efficiency effects from interference management, focusing the analysis on relay selection and hardware-aware energy modeling.

The BS and relays use multiple transmit antennas while UEs have a single receive antenna. The number of effective transmit chains at each relay sets its serving capacity, limiting the number of users it can simultaneously support. All relays operate in half-duplex mode and, unless otherwise stated, employ decode-and-forward operation, regenerating a clean copy of the signal before forwarding.

The system operates in the mmWave regime. Although mmWave links incur higher free-space path loss than sub-6~GHz systems, the higher achievable antenna gains can offset much of this loss; regardless of frequency, the WF framework enables a true energy-consumption analysis while highlighting the role of relay-assisted transmission for coverage and energy efficiency. A minimum received power constraint is adopted as a proxy for communication feasibility, which is appropriate in mmWave or other directional-beamforming systems where link availability is governed by link-budget limitations rather than interference.

Large-scale path loss is modeled deterministically as a function of distance, consistent with dominant line-of-sight or quasi-line-of-sight mmWave propagation; small-scale fading is not considered, and noise is additive white Gaussian (AWGN) with power spectral density $N_0$. The transmit powers of the BS and relays are fixed and may differ across nodes. Relays are heterogeneous in transmit power, antenna gain, non-path power consumption, and waste-factor parameters~\cite{sevim2025theoretical}, enabling realistic modeling of practical deployments and playing a central role in the energy-aware assignment strategies developed here.

\begin{figure}
    \centering
    \includegraphics[width=0.3\linewidth]{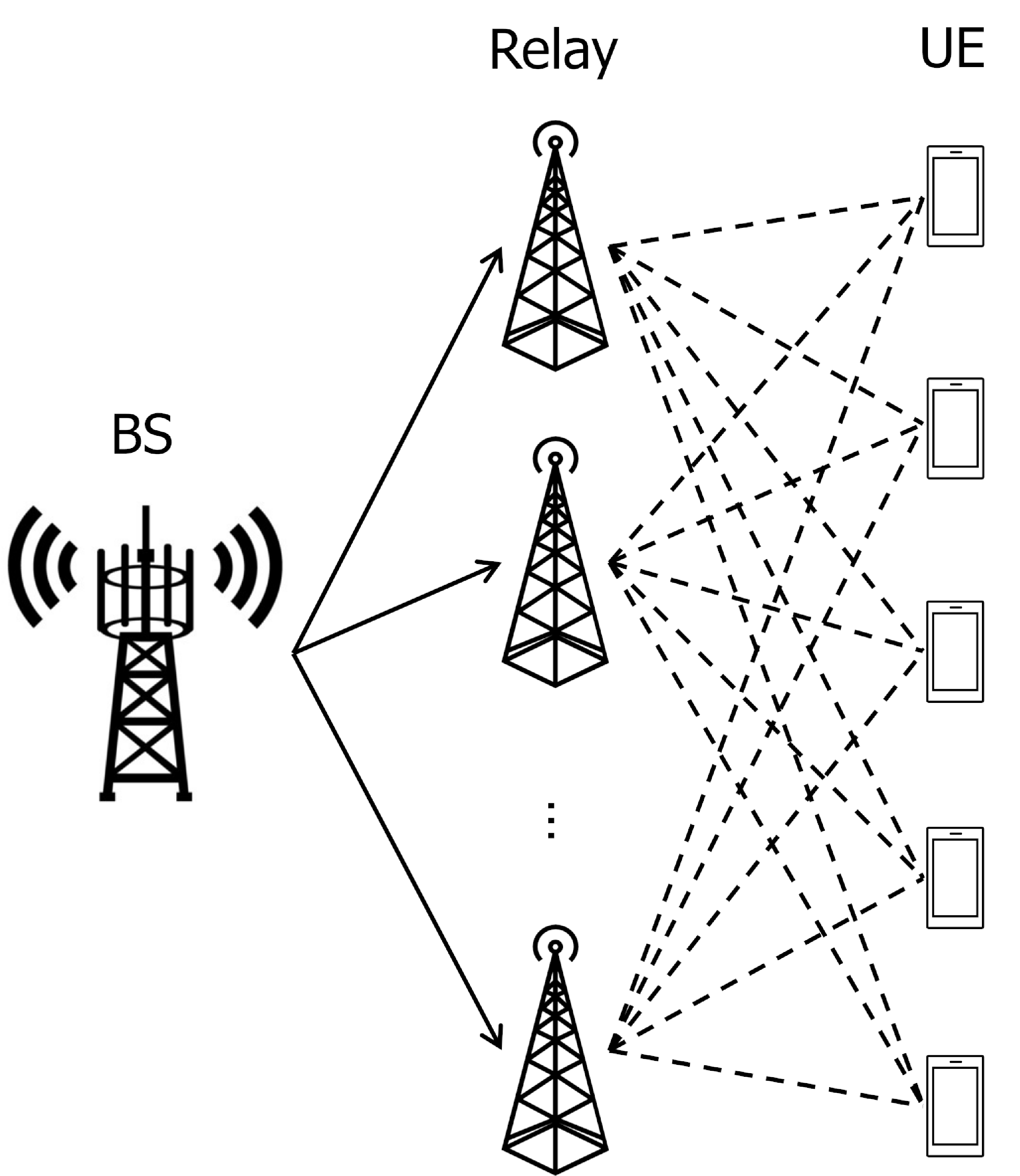}
    \caption{Illustration of the wireless system model considered in this paper. Direct BS–UE links are omitted to avoid visual clutter for the reader.}
    \label{fig:placeholder}
\end{figure}

Table~\ref{tab:symbols} summarizes the notation, and Fig.~\ref{fig:placeholder} illustrates the system model.

\section{Problem Formulation}
\label{sec:problem-form}

The linear channel gains for BS$\rightarrow$relay $r$, relay $r\rightarrow$UE $k$, and BS$\rightarrow$UE $k$ are
$G_{\mathrm{BS},r}$, $G_{r,k}$, and $G_{\mathrm{BS},k}$, respectively, incorporating path loss gains.  
The corresponding Shannon capacities are
\begin{align}
C_{\mathrm{BS},r} &= B \log_2\!\left(1 + \frac{P_{tx}^{BS} G_{tx}^{BS}G_{\mathrm{BS},r}}{N_0 B}\right), \\
C_{r,k} &= B \log_2\!\left(1 + \frac{P_{tx}^{r} G_{tx}^{r}G_{r,k}}{N_0 B}\right), \\
C_{\mathrm{BS},k} &= B \log_2\!\left(1 + \frac{P_{tx}^{BS} G_{tx}^{BS}G_{\mathrm{BS},k}}{N_0 B}\right).
\end{align}

DF relays are considered here because they regenerate a clean copy of the transmitted signal, preventing noise amplification.

For each UE $k$, define a binary vector
\[
\mathbf{x}_k = [x_{k,1},\ldots,x_{k,R}], \qquad x_{k,r}\in\{0,1\},
\]
where $x_{k,r}=1$ indicates that relay $r$ serves UE $k$.  
Direct BS service is represented by a binary variable $s_k\in\{0,1\}$.

The aggregate received power at UE $k$ is
\begin{equation}
P^{\mathrm{rx}}_k
= \sum_{r=1}^{R} x_{k,r} P_{tx}^r G_{tx}^r G_{r,k}
+ s_k P_{tx}^{BS} G_{tx}^{BS} G_{\mathrm{BS},k},
\label{eq:p_rx}
\end{equation}
with a minimum QoS constraint
\begin{equation}
P^{\mathrm{rx}}_k \ge P_{\min}.
\end{equation}

Relays can have a user-serving limit
\begin{equation}
\sum_{k=1}^{K} x_{k,r} \le L_r,
\end{equation}
where $L_r$ is the maximum number of UEs that relay $r$ can simultaneously support.  
Physically, the limit $L_r$ corresponds to the number of effective transmit antennas or parallel transmission chains available at relay $r$.

In the first case, each user is served by exactly one node, either a relay or the base station, i.e., either a single relay or the direct BS link.

\begin{equation}
    \sum_{r=1}^{R} x_{k,r} + s_k = 1.
\end{equation}

\begin{figure*}[t!]
    \centering
    \begin{subfigure}[b]{0.45\textwidth}
        \centering
        \includegraphics[width=\linewidth]{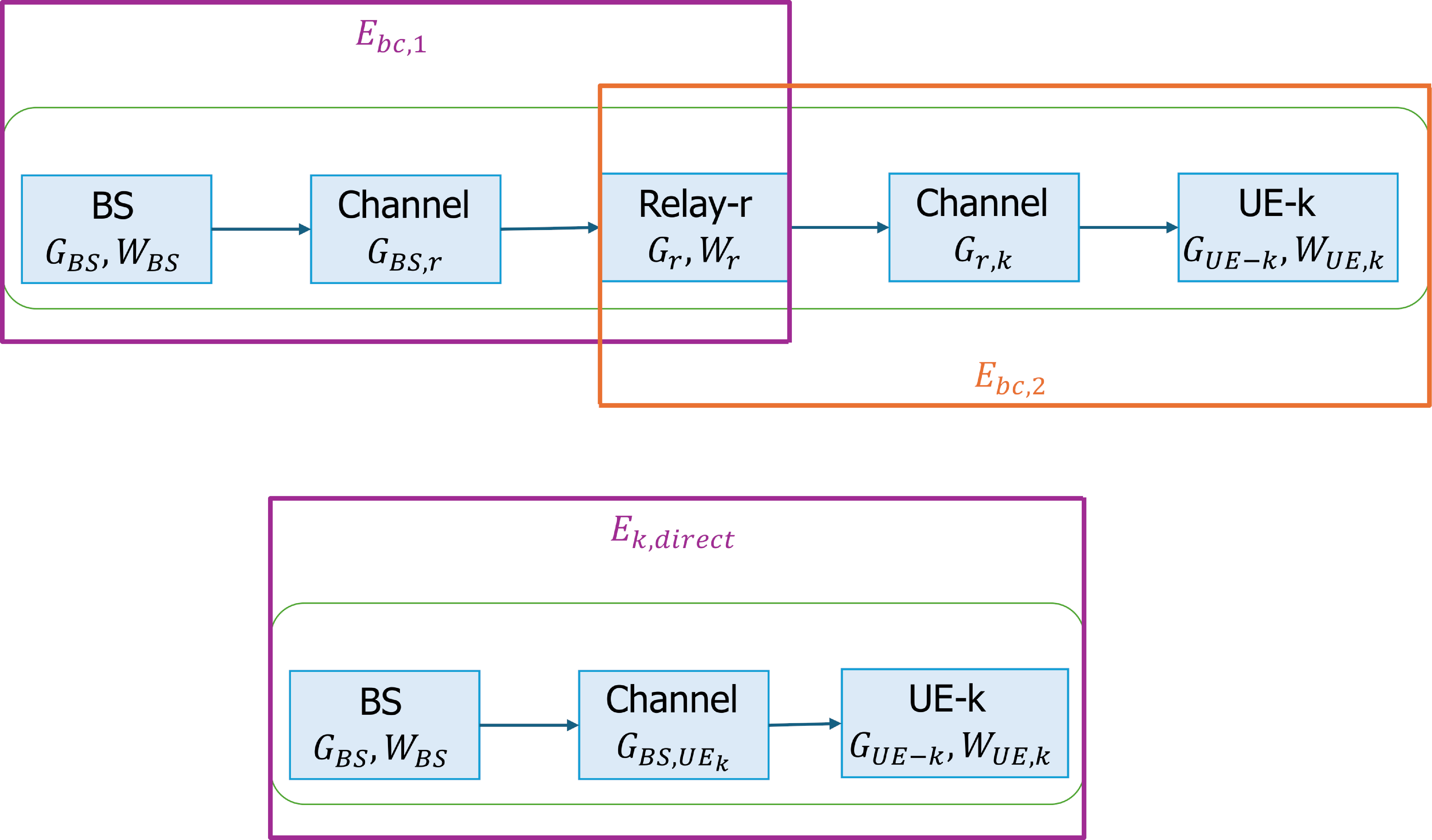}
    \caption{Case 1}
    \label{fig:case1}
    \end{subfigure}%
    ~ 
    \begin{subfigure}[b]{0.45\textwidth}
        \centering
        \includegraphics[width=\linewidth]{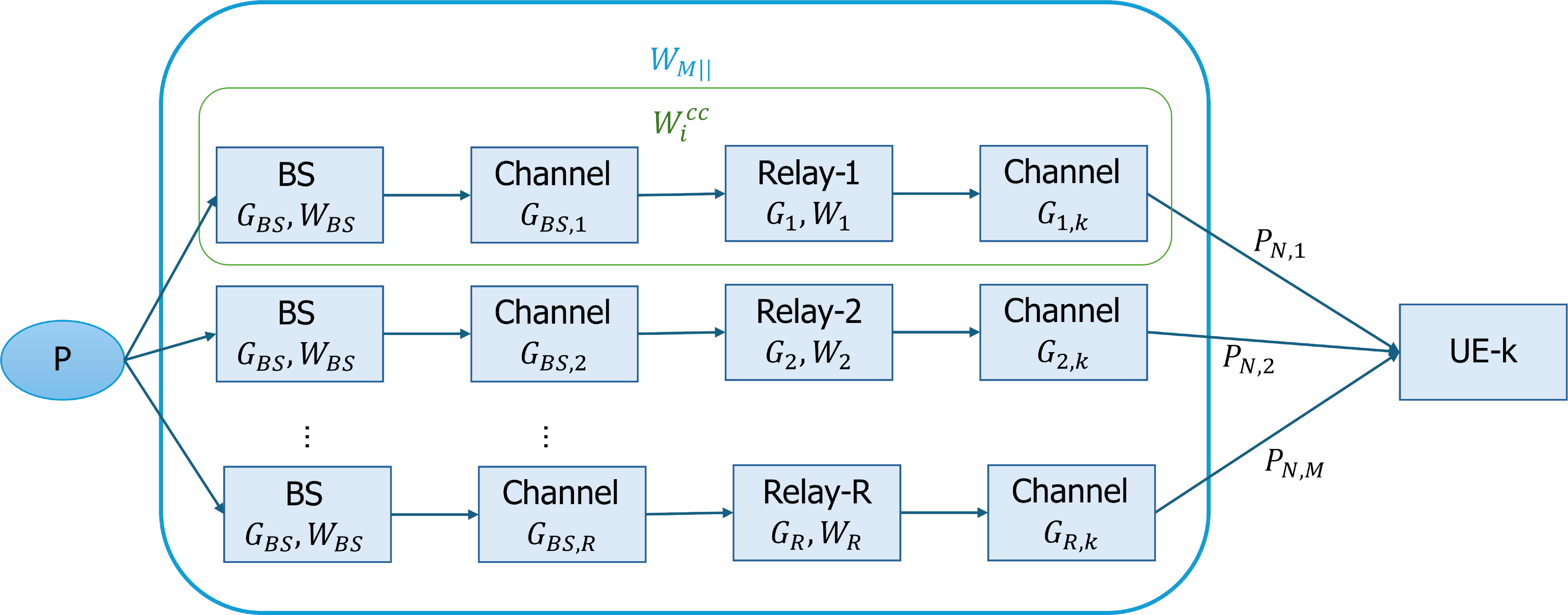}
    \caption{Case 2}
    \label{fig:case2}
    \end{subfigure}
    \caption{Graphical representation of the energy-per-bit computation for the two relay assignment cases considered in this paper. In the single-relay case, each user is served by at most one relay (or directly by the base station), resulting in a non-parallel cascaded transmission path. In contrast, in the multi-relay case, a user can be simultaneously served by multiple relays, leading to a parallel cascaded system structure.}
    \label{fig:case1-2}
\end{figure*}

The objective is to minimize the energy required per bit transfer. Assuming $E_{k,direct}$ is the energy consumption for the direct link, the objective function for this case can be expressed as:

\begin{equation}
\begin{aligned}
&\min_{x_{k,r}} \quad \sum_k \sum_r x_{k,r} E_{bc}^k(r) + s_kE_{k,direct},\\
&\textrm{s.t.} \quad \sum_{r=1}^{R} x_{k,r} + s_k = 1.
\end{aligned}
\label{eq:case1obj}
\end{equation}

Fig.~\ref{fig:case1} provides an illustration of such a scenario. Here, when expressing the total energy required to transfer a bit, the WF-based expression from \cite{sevim2025theoretical} is utilized.

\begin{align}
    &E_{bc}^k(r) = E_{bc,1}(r)+E_{bc,2}(r) \\
    &\begin{aligned}
        E_{bc,1}(r)&= \frac{P_{NP}^1}{C_1} + \ln(2) N_0 W_1 = \frac{P_{NP}^1}{C_1} + \ln(2) N_0 \frac{W_{BS}}{G_{rx}^r G_{BS,r}}, \\
    \end{aligned}\\
    &\begin{aligned}
       E_{bc,2}(r) & = \frac{P_{NP}^2}{C_2} + \ln(2) N_0 W_2 = \frac{P_{NP}^2}{C_2} + \ln(2) N_0 \frac{W_r}{G_{rx}^k G_{r,k}}, \\
    \end{aligned}\\
    &\begin{aligned}
        E_{bc}^k(r) = \frac{P_{NP}^1}{C_1} + \ln(2) N_0 \frac{W_{BS}}{G_{rx}^r G_{BS,r}} + \frac{P_{NP}^2}{C_2}
 + \ln(2) N_0 \frac{W_r}{G_{rx}^k G_{r,k}}.
    \end{aligned}   
\end{align}

The non-path power consumption of the first hop $P_{NP}^1$ consists of the non-path power consumption of the base station and relay, while the non-path power consumption of the second hop $P_{NP}^2$ consists of the non-path power consumption of the relay and the user end device. Thus:

\begin{equation}
    \begin{aligned}
    E_{bc}^k(r) &=  \frac{P_{NP}^{BS} + P_{NP}^r}{C_{BS,r}} + \frac{P_{NP}^{k} + P_{NP}^r}{C_{r,k}} 
     + \ln(2) N_0 \left(\frac{W_{BS}}{G_{rx}^r G_{BS,r}} + \frac{W_r}{G_{rx}^k G_{r,k}}\right) . 
\end{aligned}
\label{eq:case1ek}
\end{equation}

And the energy consumption of the direct link can simply be written as:

\begin{equation}
    \begin{aligned}
        E_{k,direct}&= \frac{P_{NP}^{BS} + P_{NP}^{k}}{C_{BS,k}} + \ln(2) N_0 W_{direct} 
        = \frac{P_{NP}^{BS} + P_{NP}^{k}}{C_{BS,k}} + \ln(2) N_0 \frac{W_{BS}}{G_{rx}^k G_{BS,k}}. 
    \end{aligned}
    \label{eq:case1edirect}
\end{equation}

Now Eqs.~\eqref{eq:case1ek} and \eqref{eq:case1edirect} can be written into \eqref{eq:case1obj} to obtain the optimization formula.

First, a structured special case is considered in which the relays have no limit on the number of users they can serve. Although the assumption of unlimited relay capacity represents an idealized scenario, it will help build up the problem. Mathematically, the case where the relay can serve any number of users can be expressed as follows:

\begin{equation}
    \sum_{k=1}^K x_{k,r} \le K.
\end{equation}

With unlimited relay capacity, each user selects its serving node independently, so the problem decouples and is solved optimally by the greedy assignment of Section~\ref{sec:methodology}.

In the capacity-constrained setting, each relay is limited to serving a finite number of users. The physical interpretation of this limitation is that each relay has a finite number of transmit antennas (RF chains). Consequently, there is an upper bound on the number of orthogonal transmission links that a relay can simultaneously maintain.
\begin{equation}
    \sum_{k=1}^{K} x_{k,r} \le L_r.
\end{equation}

The relay-capacity-constrained assignment scenario introduces combinatorial coupling, and its optimization falls within the binary linear programming domain. As discussed later in Section~\ref{sec:methodology}, this scenario can be formulated as a \textbf{Min-Cost Flow} problem and can be solved exactly using well-known linear programming approaches, such as successive shortest path or cost-scaling methods.

In the second case, each user can be simultaneously served by multiple relays and/or the BS:

\begin{equation}
    0<\sum_{r=1}^{R} x_{k,r} + s_k \le R + 1.
\end{equation}

The objective is to minimize the energy required per bit transfer. The objective function for this case can be expressed as:

\begin{equation}
\begin{aligned}
&\min_{x_{k,r}} \quad \sum_k E_{bc}^k,\\
&\textrm{s.t.} \hspace{9pt} 0<\sum_{r=1}^{R} x_{k,r} + s_k \le R + 1.\\
&P_{\min} \le P_{rx,k}, \quad k\in{0,1,...,K}.
\end{aligned}
\label{eq:case2obj-1}
\end{equation}

Fig.~\ref{fig:case2} provides an illustration of such a scenario. Here, when expressing the total energy required to transfer a bit, the WF-based expression from \cite{sevim2025theoretical} is utilized. The WF for parallel cascaded systems is studied and derived in \cite{ying2024using}. The WF formulation for non-coherent combining of parallel systems is used. According to \cite{ying2024using}, the WF for a non-coherently combined M parallel cascaded system (as illustrated in Fig.~\ref{fig:miso_wf}) is given as:

\begin{figure}
    \centering
    \includegraphics[width=.4\linewidth]{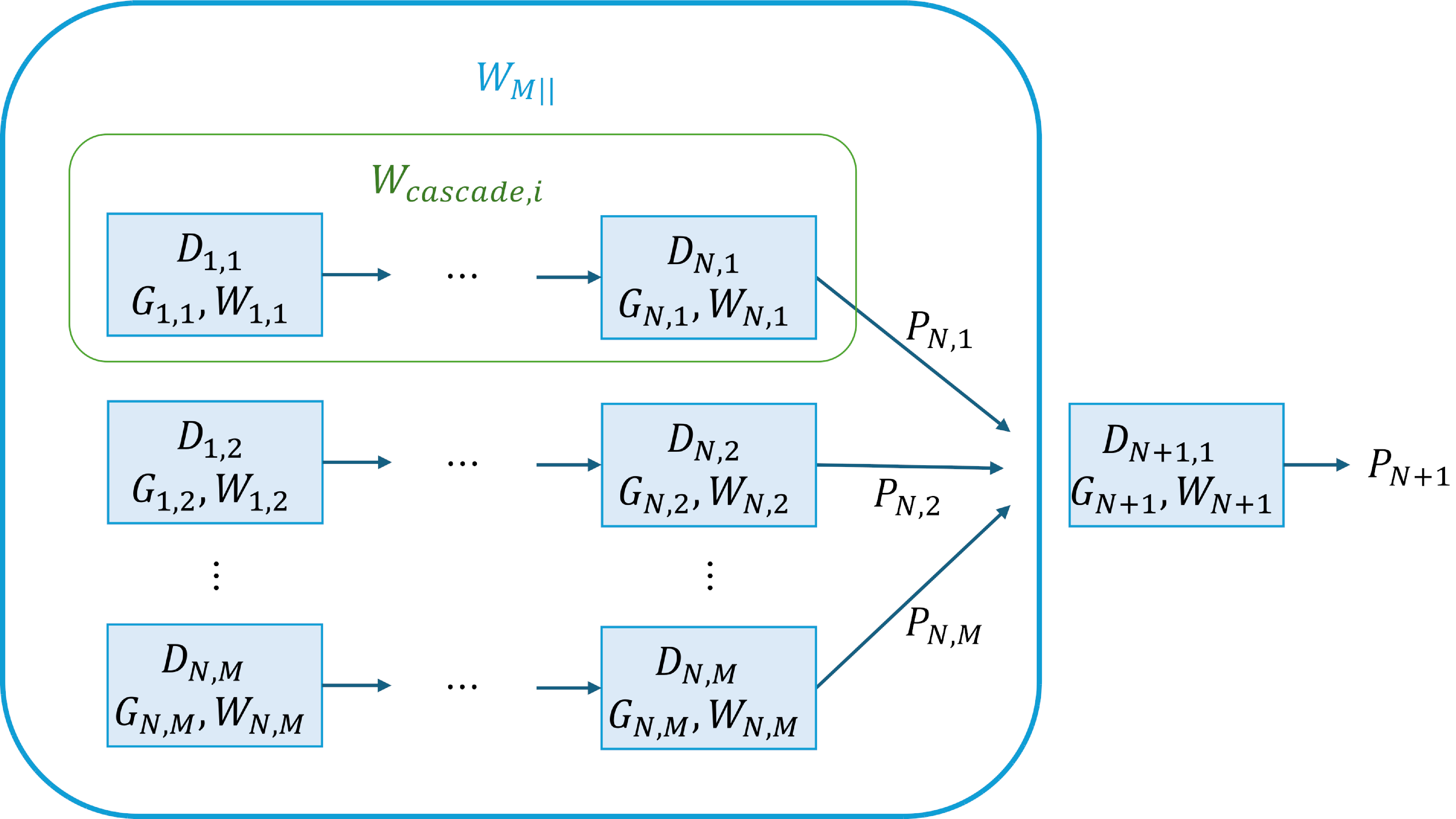}
    \caption{The figure illustrates a multi-stage cascaded transmission structure in which multiple parallel device chains contribute to a final multi-input single-output (MISO) link. Each device block is characterized by its gain and waste factor, and the overall cascade aggregates both propagation-dependent gains and signal-independent hardware losses \cite{ying2023waste}.}
    \label{fig:miso_wf}
\end{figure}

\begin{equation}
    \begin{aligned}
        W_{M||} = \frac{\sum_{i=1}^M \left( P_i \times W_{cascade,i} \right)}{\sum_{i=1}^M P_i} = \frac{\sum_{i=1}^M \left( \gamma_i \times W_{cascade,i} \right)}{\sum_{i=1}^M \gamma_i},
    \end{aligned}
\label{eq:misowf}
\end{equation}
where $[P_1,P_2,...,P_M] = P[\gamma_1,\gamma_2,...,\gamma_M]$, where $P$ is a constant base power and $\gamma_i$ are scaling factors that illustrate the relative contribution to the total received power from each cascade output. 

Non-coherent combining is considered as it does not require phase synchronization across distributed relays, making it more practical for decentralized and heterogeneous deployments.

In the case depicted in Fig.~\ref{fig:case2}, the received power contributions can be expressed as \[ [P_1,P_2,...,P_M] = P[G_1 G_{1,k},G_2 G_{2,k},...,G_M G_{M,k}].\] 

Then, the overall WF for the non-coherently combined cascaded system at user $k$ becomes:

\begin{equation}
    W_k^{nc} = \frac{\sum_{r=1}^Rx_{k,r} G_r G_{r,k} W_{r}^{cc}}{\sum_{r=1}^Rx_{k,r} G_r G_{r,k}},
\label{eq:misowf-1}
\end{equation}
where, $W_{r}^{cc}$ is the cascaded link through relay $r$ and is calculated as, by Eq.~\eqref{eq:waste_factor_cascade}:
\begin{equation}
    W_{r}^{cc} = \frac{1}{G_{r,k}}+\frac{W_r-1}{G_{r,k}}=\frac{W_r}{G_{r,k}}.
    \label{eq:caswf}
\end{equation}

Replacing Eq.~\eqref{eq:caswf} into Eq.~\eqref{eq:misowf-1} yields:
\begin{equation}
    \begin{aligned}
        W_k^{nc} = \frac{\sum_{r=1}^Rx_{k,r} G_r G_{r,k} W_{r}^{cc}}{\sum_{r=1}^Rx_{k,r} G_r G_{r,k}} = \frac{\sum_{r=1}^Rx_{k,r} G_r G_{r,k} \frac{W_r}{G_{r,k}}}{\sum_{r=1}^Rx_{k,r} G_r G_{r,k}}
        =\frac{\sum_{r=1}^Rx_{k,r} G_r W_r}{\sum_{r=1}^Rx_{k,r} G_r G_{r,k}}.
    \end{aligned}
\end{equation}

Then, the WF of the overall multi-relay system for a single user can be obtained as follows:

\begin{equation}
    \begin{aligned}
    W^k = W_{UE_k}+\frac{W_k^{nc}-1}{G_{UE_k}}=W_{UE_k}+\frac{\frac{\sum_{r=1}^Rx_{k,r} G_r W_r}{\sum_{r=1}^Rx_{k,r} G_r G_{r,k}}-1}{G_{UE_k}}
    =W_{UE_k}+\frac{\sum_{r=1}^Rx_{k,r} G_r (W_r-G_{r,k})}{\sum_{r=1}^Rx_{k,r} G_{UE_k} G_r G_{r,k}}.
\end{aligned}
\end{equation}

Now, the expression for the total energy-per-bit for user $k$ can be obtained as:
\begin{equation}
    \begin{aligned}
        E_{bc}^k &= \sum_r x_{k,r} \left( \frac{P_{NP}^{BS}+P_{NP}^{r}}{C_{BS,r}} +\ln(2)N_0 \left( \frac{W_{BS}}{G_{BS,r}G_{rx}^r} \right) \right)\\
        &+\sum_r x_{k,r} \frac{P_{NP}^r}{C_{r,k}}+\frac{P_{NP}^{UE_k}}{\sum_r x_{k,r}C_{r,k}} \\
        &+ \ln(2) N_0 \left( W_{UE_k}+\frac{\sum_{r=1}^Rx_{k,r} G_r (W_r-G_{r,k})}{\sum_{r=1}^Rx_{k,r} G_{UE_k} G_r G_{r,k}} \right).
    \end{aligned}
    \label{eq:case2-ebc}
\end{equation}

Then, the final objective for \textbf{Case 2} can be expressed as:
\begin{equation}
\begin{aligned}
&\min_{x_{k,r}} \quad \sum_k E_{bc}^k,\\
&\textrm{s.t.~} 0 < \sum_r x_{k,r} , x_{k,r}\in{0,1},\\
&P_{\min} \le P_{rx,k}, \quad k\in{0,1,...,K}.
\end{aligned}
\label{eq:case2obj}
\end{equation}
where $E_{bc}^k$ is given in Eq.~\eqref{eq:case2-ebc} and $P_{rx,k}$ can be obtained by Eq.~\eqref{eq:p_rx}.

The problem in Eq.~\eqref{eq:case2obj} is highly non-linear and non-convex, so finding the exact solution for large $K$ and $R$ becomes infeasible. Therefore, in this paper, some sub-optimal solutions are proposed which have acceptable complexity and are compared with the exhaustive search solution when possible (for small $K$ and $R$).

\begin{figure*}[t]
    \centering
    \includegraphics[width=0.7\linewidth]{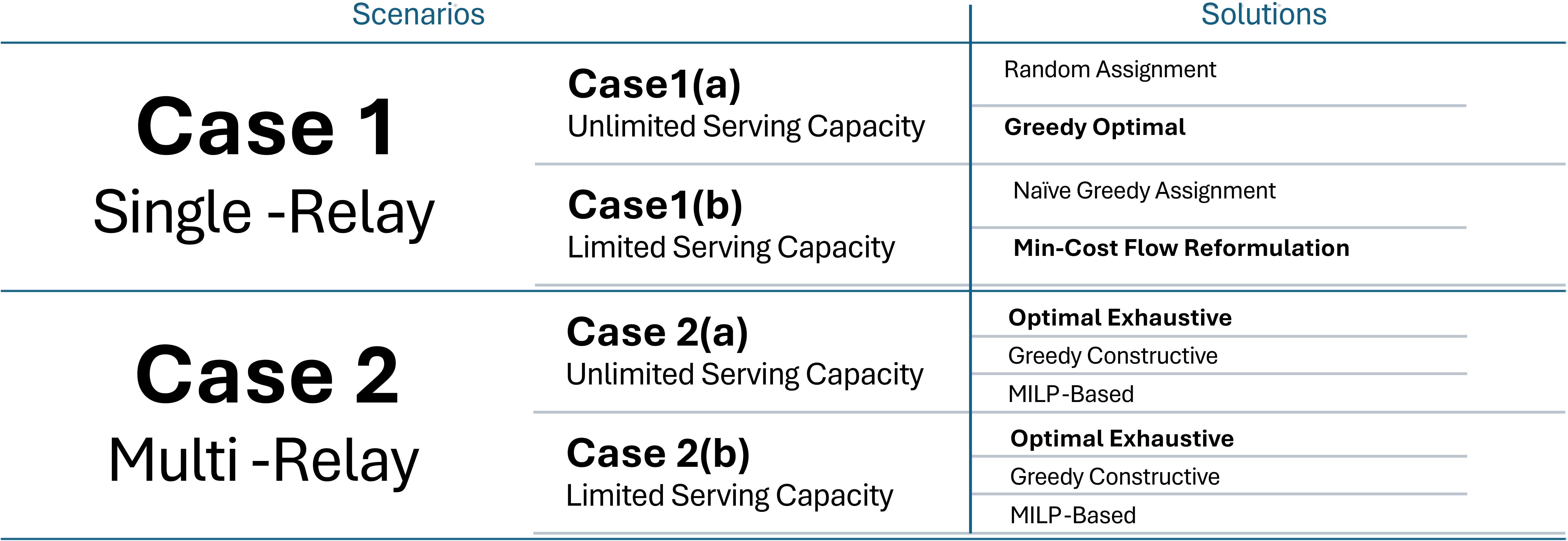}
    \caption{Graphical summary of the four relay-assignment scenarios formulated in Section~\ref{sec:problem-form} and the corresponding solution strategies developed in Section~\ref{sec:methodology}. Bold fonts indicate exact formulations.}
    \label{fig:case_chart}
\end{figure*}

Fig.~\ref{fig:case_chart} provides a consolidated overview of the four relay-assignment scenarios considered in this work and the corresponding solution methodologies proposed for each case, which will be explained in Section \ref{sec:methodology}. The classification follows the structural distinctions introduced in Section~\ref{sec:problem-form}, namely (i) single-relay versus multi-relay serving and (ii) absence or presence of relay capacity constraints. 

\section{Methodology}
\label{sec:methodology}
This section presents algorithmic solutions for the relay-assignment problems of Section~\ref{sec:problem-form}. Case~1(a) decomposes across users and admits a globally optimal greedy solution (Algorithm~1). Case~1(b) couples users through relay capacity but remains exactly solvable via a min-cost flow reformulation (Algorithm~2), with a capacity-aware greedy baseline for comparison (Algorithm~3). The multi-relay case (Case~2) is generally intractable; it is addressed with a constructive greedy approximation (Algorithm~4) and a MILP formulation over restricted candidate sets (Algorithm~5), with exhaustive search as a small-scale benchmark.

\subsection{Single-Relay Assignment}
For the optimization problem of \textbf{Case~1}, two scenarios are considered. 
In the first scenario, relays are assumed to have unlimited serving capability, i.e., there is no restriction on the number of users that can be simultaneously supported by a relay. 
Furthermore, no received power constraint is imposed in this case. 
Since users are assigned orthogonal frequency bands, inter-user interference is neglected.

The total energy-per-bit minimization problem is:
\begin{equation}
\min_{\{x_{k,r}\}} \sum_{k=1}^{K} E_{bc}^k(\mathbf{x}_k),
\end{equation}
subject to
\begin{equation}
\sum_{r \in \mathcal{R} \cup \{\mathrm{BS}\}} x_{k,r} = 1, 
\quad x_{k,r} \in \{0,1\},
\end{equation}
decomposes into $K$ independent subproblems of the form
\begin{equation}
\min_{r \in \mathcal{R} \cup \{\mathrm{BS}\}} E_{bc}^k(r), \quad \forall k.
\end{equation}

Formally, the greedy assignment is given by
\begin{equation}
x_{k,r_k^\star}=1,r_k^{\star} = \arg\min_{r \in \mathcal{R} \cup \{\mathrm{BS}\}} E_{bc}^k(r),
\end{equation}
which is globally optimal under the stated assumptions due to the absence of coupling constraints.

\begin{algorithm}[t]
\caption{Greedy Single-Link Assignment for Case~1}
\label{alg:greedy_single}
\begin{algorithmic}[1]
\REQUIRE Set of users $\mathcal{K}$, relays $\mathcal{R}$, energy-per-bit values $\{E_{bc}^k(r)\}$ for all $k \in \mathcal{K}$ and $r \in \mathcal{R} \cup \{\mathrm{BS}\}$
\ENSURE Serving node assignment $\{r_k^\star\}$ for all users

\FOR{each user $k \in \mathcal{K}$}
    \STATE Compute $E_{bc}^k(r)$ for all $r \in \mathcal{R} \cup \{\mathrm{BS}\}$
    \STATE Select
    \[
    r_k^\star \gets \arg\min_{r \in \mathcal{R} \cup \{\mathrm{BS}\}} E_{bc}^k(r)
    \]
\ENDFOR
\RETURN $\{r_k^\star\}_{k \in \mathcal{K}}$
\end{algorithmic}
\end{algorithm}

Algorithm~\ref{alg:greedy_single} yields the globally optimal solution for Case~1 due to the absence of coupling constraints among users and serves as a baseline for more constrained scenarios considered in subsequent sections.

In the second scenario, each user $k \in \mathcal{K}$ is served by \emph{at most one} relay or directly by the base station (BS). Unlike Case~1(a), each relay has a finite serving capacity, which limits the number of users it can simultaneously support. This capacity can be physically interpreted as the number of available transmit antennas or RF chains at the relay.

Let $\mathcal{R}=\{1,\dots,R\}$ denote the set of relays and let $\mathrm{BS}$ represent the direct link option. Define the binary decision variable
\[
x_{k,r} =
\begin{cases}
1, & \text{if user } k \text{ is served by node } r,\\
0, & \text{otherwise},
\end{cases}
\quad r \in \mathcal{R} \cup \{\mathrm{BS}\}.
\]

The energy-per-bit cost incurred by serving user $k$ through node $r$ is denoted by $E_{bc}^k(r)$.

The optimization problem is given by
\begin{equation}
\begin{aligned}
    \min_{\{x_{k,r}\}} \quad 
& \sum_{k\in\mathcal{K}} \sum_{r\in\mathcal{R}\cup\{\mathrm{BS}\}} 
E_{bc}^k(r)\, x_{k,r} \\
\text{s.t.} \quad
& \sum_{r\in\mathcal{R}\cup\{\mathrm{BS}\}} x_{k,r} = 1,
&& \forall k \in \mathcal{K}, \\
& \sum_{k\in\mathcal{K}} x_{k,r} \le L_r,
&& \forall r \in \mathcal{R}, \\
& x_{k,r} \in \{0,1\},
&& \forall k,r.
\end{aligned}
 \label{eq:case1b_obj}
\end{equation}

Here, $L_r$ denotes the maximum number of users that relay $r$ can serve. The BS is assumed to have unlimited capacity and therefore does not require a capacity constraint.

\subsubsection{Solving by Min-Cost Flow Reformulation}

Problem~\eqref{eq:case1b_obj} can be equivalently reformulated as a \emph{minimum-cost flow (MCF)} problem on a directed graph $\mathcal{G}=(\mathcal{V},\mathcal{E})$, constructed as follows:

\begin{itemize}
    \item A source node $s$ and a sink node $t$.
    \item A node for each user $k \in \mathcal{K}$.
    \item A node for each relay $r \in \mathcal{R}$ and one node for the BS.
\end{itemize}

The edge set $\mathcal{E}$ is defined by:
\begin{itemize}
    \item Edges $(s \rightarrow k)$ with capacity $1$ and cost $0$, $\forall k \in \mathcal{K}$.
    \item Edges $(k \rightarrow r)$ with capacity $1$ and cost $E_{bc}^k(r)$, $\forall k \in \mathcal{K}, r \in \mathcal{R} \cup \{\mathrm{BS}\}$.
    \item Edges $(r \rightarrow t)$ with capacity $L_r$ and cost $0$, $\forall r \in \mathcal{R}$.
    \item Edge $(\mathrm{BS} \rightarrow t)$ with infinite capacity and cost $0$.
\end{itemize}

Each unit of flow corresponds to the assignment of exactly one user to a serving node. The relay capacity constraints are naturally enforced through the edge capacities $(r \rightarrow t)$.

The resulting flow network is integral and totally unimodular; therefore, the optimal solution to the min-cost flow problem yields integer flows without the need for additional integrality constraints. Consequently, the globally optimal assignment can be obtained in polynomial time using standard min-cost flow algorithms, such as successive shortest-path or cost-scaling methods.

The min-cost flow formulation in \eqref{eq:case1b_obj} enables efficient and scalable optimization, rigorously captures relay capacity limitations, and serves as a benchmark against heuristic strategies.

\begin{algorithm}[t]
\caption{Min-Cost Flow Solver for Case 1(b)}
\label{alg:mcf}
\begin{algorithmic}[1]
\REQUIRE User set $\mathcal{K}$, relay set $\mathcal{R}$, relay capacities $\{L_r\}$,
energy costs $\{E_{bc}^k(r)\}$
\ENSURE Optimal assignment $\{x_{k,r}^\star\}$

\STATE Construct directed graph $\mathcal{G}=(\mathcal{V},\mathcal{E})$:
\STATE \quad Add source $s$ and sink $t$
\STATE \quad Add user nodes $k \in \mathcal{K}$
\STATE \quad Add relay nodes $r \in \mathcal{R}$ and BS node

\FORALL{$k \in \mathcal{K}$}
    \STATE Add edge $(s \rightarrow k)$ with capacity $1$ and cost $0$
    \FORALL{$r \in \mathcal{R} \cup \{\mathrm{BS}\}$}
        \STATE Add edge $(k \rightarrow r)$ with capacity $1$ and cost $E_{bc}^k(r)$
    \ENDFOR
\ENDFOR

\FORALL{$r \in \mathcal{R}$}
    \STATE Add edge $(r \rightarrow t)$ with capacity $L_r$ and cost $0$
\ENDFOR
\STATE Add edge $(\mathrm{BS} \rightarrow t)$ with infinite capacity and cost $0$

\STATE Solve the minimum-cost flow problem with total flow $|\mathcal{K}|$
\STATE Extract assignments:
\[
x_{k,r}^\star =
\begin{cases}
1, & \text{if flow passes through edge } (k \rightarrow r),\\
0, & \text{otherwise}
\end{cases}
\]

\STATE \RETURN $\{x_{k,r}^\star\}$
\end{algorithmic}
\end{algorithm}

\subsubsection{Naïve Greedy Baseline}

As a low-complexity baseline for \textbf{Case~1(b)}, a naïve capacity-aware greedy heuristic is also considered. Operating on the energy-per-bit cost matrix $E_{bc}\in\mathbb{R}^{K\times(R+1)}$, it processes users sequentially: for each user $k$, candidate nodes in $\mathcal{R}_0=\mathcal{R}\cup\{\mathrm{BS}\}$ are sorted by ascending $E_{bc}^k(r)$, and the lowest-cost feasible option is selected. The BS is always accepted (assumed to have ample capacity); a relay is accepted only if it has remaining capacity. Assignments are fixed once made, with no backtracking, and the relay load is updated immediately. The procedure is summarized in Algorithm~\ref{alg:case1b_naive}, with total cost $E_{\mathrm{tot}}^{\text{naive}} = \sum_{k\in\mathcal{K}} E_{bc}^k(r_k)$.

\begin{algorithm}[t]
\caption{Naïve Capacity-Aware Assignment for Case~1(b)}
\label{alg:case1b_naive}
\begin{algorithmic}[1]
\REQUIRE Energy costs $\{E_{bc}^k(r)\}$, relay capacities $\{L_r\}_{r\in\mathcal{R}}$
\ENSURE Assignment $\{r_k\}$ and total energy cost

\STATE Initialize relay loads $L_r^{\mathrm{used}} \gets 0$ for all $r\in\mathcal{R}$
\FOR{each user $k=1,\ldots,K$}
    \STATE Sort $r\in\mathcal{R}_0$ in ascending order of $E_{bc}^k(r)$
    \FOR{each $r$ in sorted order}
        \IF{$r=\mathrm{BS}$}
            \STATE Assign $r_k \gets \mathrm{BS}$; \textbf{break}
        \ELSIF{$L_r^{\mathrm{used}} < L_r$}
            \STATE Assign $r_k \gets r$
            \STATE $L_r^{\mathrm{used}} \gets L_r^{\mathrm{used}} + 1$
            \STATE \textbf{break}
        \ENDIF
    \ENDFOR
\ENDFOR
\end{algorithmic}
\end{algorithm}

Its complexity is $\mathcal{O}(K(R+1)\log(R+1))$, making it highly scalable, but the sequential, irrevocable assignments are sensitive to user ordering: early users may occupy relays that would yield larger global savings elsewhere. It therefore serves only as a low-complexity reference and is not expected to be optimal.

Restricting each user to a single transmitter enables efficient optimization but ignores the spatial diversity of dense relay deployments, where several relays can jointly raise received power to meet QoS or improve efficiency. This motivates Case~2, where multi-relay cooperation introduces combinatorial coupling and nonlinear energy interactions, requiring the strategies developed next.

\subsection{Multi-Relay Assignment}
\label{subsec:case2_methods}

In \textbf{Case~2}, a user can be simultaneously served by multiple relays and/or directly by the base station (BS). The objective is to minimize the energy per bit while satisfying a minimum received power constraint. Section~\ref{sec:methodology} presents algorithmic solutions used in this work: a constructive greedy approximation and an optimization-based mixed-integer linear programming (MILP) formulation.

\subsubsection{Constructive Approximation of the Exhaustive Search}

Let $\mathcal{R}=\{1,\ldots,R\}$ denote the set of relays and define the augmented transmitter set
$\mathcal{R}_0=\mathcal{R}\cup\{\mathrm{BS}\}$.
For a given user $k$, let $\mathcal{S}_k\subseteq\mathcal{R}_0$ denote the serving set.
The received power at user $k$ is given by
\begin{equation}
P_k^{\mathrm{rx}}(\mathcal{S}_k)
=
\sum_{r\in\mathcal{S}_k} P_{k,r}^{\mathrm{rx}},
\end{equation}
where $P_{k,r}^{\mathrm{rx}}$ denotes the received power contribution from transmitter $r$.
A user is considered feasible only if
\begin{equation}
P_k^{\mathrm{rx}}(\mathcal{S}_k) \ge P_{\min}.
\end{equation}

The energy per bit associated with serving set $\mathcal{S}_k$ is denoted by
$E_{bc}^k(\mathcal{S}_k)$ and is computed using the exact multi-relay energy model introduced in
Section~\ref{sec:system_model}.
In the absence of relay capacity limits, the optimization problem for user~$k$ can be written as
\begin{equation}
\min_{\mathcal{S}_k\subseteq\mathcal{R}_0}
E_{bc}^k(\mathcal{S}_k)
\quad
\text{s.t.}
\quad
P_k^{\mathrm{rx}}(\mathcal{S}_k)\ge P_{\min},
\end{equation}
which is combinatorial and exponential in $R$.
When relay capacity limits are imposed, the serving sets of different users become coupled through
shared relay resources.

A \emph{constructive greedy algorithm} approximates the exhaustive search by incrementally building each user's serving set: starting from the empty set, it adds one transmitter per iteration until $P_{\min}$ is met. If no single available transmitter closes the remaining power deficit, the one with the largest received-power contribution is added; otherwise the one yielding the minimum incremental energy per bit under $E_{bc}^k(\cdot)$ is chosen. Capacity is enforced via a global per-relay usage counter, an exhausted relay is removed from all users' candidate sets, and users with no feasible set are marked infeasible.

\begin{algorithm}[h]
\caption{Capacity-Aware Constructive Multi-Relay Selection}
\label{alg:constructive}
\begin{algorithmic}[1]

\REQUIRE User set $\mathcal{K}$, relay set $\mathcal{R}$,
minimum received power $P_{\min}$,
relay capacities $\{L_r\}_{r\in\mathcal{R}}$,
energy-per-bit evaluation routine $E_{bc}^k(\mathcal{S})$ for any $\mathcal{S}\subseteq\mathcal{R}_0$
\ENSURE Serving sets $\{\mathcal{S}_k\}_{k\in\mathcal{K}}$, infeasible user set $\mathcal{K}_{\mathrm{inf}}$

\STATE Initialize relay capacities $\{L_r\}_{r\in\mathcal{R}}$
\FOR{each user $k$}
    \STATE $\mathcal{S}_k \gets \emptyset$, $P_k^{\mathrm{rx}} \gets 0$
    \STATE $\mathcal{A}_k \gets \mathcal{R}_0$ \COMMENT{Available transmitters}
    \WHILE{$P_k^{\mathrm{rx}} < P_{\min}$ and $\mathcal{A}_k \neq \emptyset$}
        \IF{$\nexists r\in\mathcal{A}_k$ such that $P_{k,r}^{\mathrm{rx}} \ge P_{\min}-P_k^{\mathrm{rx}}$}
            \STATE $r^\star \gets \arg\max_{r\in\mathcal{A}_k} P_{k,r}^{\mathrm{rx}}$
        \ELSE
            \STATE $r^\star \gets \arg\min_{r\in\mathcal{A}_k} E_{bc}^k(\mathcal{S}_k\cup\{r\})$
        \ENDIF
        \STATE $\mathcal{S}_k \gets \mathcal{S}_k \cup \{r^\star\}$
        \STATE $P_k^{\mathrm{rx}} \gets P_k^{\mathrm{rx}} + P_{k,r^\star}^{\mathrm{rx}}$
        \STATE Update relay usage if $r^\star \in \mathcal{R}$
        \STATE Remove $r^\star$ from $\mathcal{A}_k$
        \STATE Remove any relay $r$ with exhausted capacity from all $\mathcal{A}_{k'}$
    \ENDWHILE
    \IF{$P_k^{\mathrm{rx}} < P_{\min}$}
        \STATE Mark user $k$ as infeasible
    \ENDIF
\ENDFOR
\end{algorithmic}
\end{algorithm}

Algorithm~\ref{alg:constructive} provides a constructive approximation of the exhaustive search solution.
Unlike relaxations or linearized formulations, it operates directly on the exact energy-per-bit expression.
In the absence of relay capacity limits, the algorithm runs independently for each user and has polynomial complexity $\mathcal{O}(R^2)$ per user.
When capacity limits are enforced, the algorithm remains polynomial but introduces coupling among users via relay availability, which can lead to suboptimal allocations relative to the global optimum.
Nevertheless, simulation results demonstrate that the proposed approach achieves performance close to the exhaustive search benchmark while maintaining scalability to realistic network sizes.

\subsubsection{MILP-Based Multi-Relay Optimization}

To obtain an optimal solution under structured candidate sets, the problem is reformulated as a mixed-integer linear program (MILP). For each user $k$, a finite candidate set
\begin{equation}
\mathcal{C}_k=\{(\mathcal{S}_k^{(i)}, E_{bc}^{k\,(i)}, P_k^{(i)})\}
\end{equation}
is generated offline by enumerating small subsets of relays and the direct BS link. Candidates violating the received power constraint, i.e., $P_k^{(i)}<P_{\min}$, are discarded.

Binary decision variables $y_{k,i}$ are introduced such that $y_{k,i}=1$ if candidate $i$ is selected for user $k$. The MILP is then formulated as

\begin{equation}
\begin{aligned}
    \min_{\{y_{k,i}\}} \quad &
\sum_{k}\sum_{i\in\mathcal{C}_k} y_{k,i} E_{bc}^{k\,(i)}, \\
\text{s.t.} \quad
& \sum_{i\in\mathcal{C}_k} y_{k,i} = 1, \quad \forall k, \\
& \sum_{k}\sum_{i\in\mathcal{C}_k} y_{k,i}\mathbbm{1}\{r\in\mathcal{S}_k^{(i)}\}
\le L_r, \quad \forall r\in\mathcal{R}, \\
& y_{k,i} \in \{0,1\}.
\end{aligned}
\end{equation}

Here, $L_r$ denotes the capacity of relay $r$, interpreted as the maximum number of users it can simultaneously serve. Since all nonlinearities are resolved during candidate generation, the resulting problem is linear in the decision variables and can be solved using standard MILP solvers.

\begin{algorithm}[h]
\caption{MILP-Based Multi-Relay Assignment}
\label{alg:milp_multirelay}
\begin{algorithmic}[1]

\REQUIRE User set $\mathcal{K}$, relay set $\mathcal{R}$,
minimum received power $P_{\min}$,
relay capacities $\{L_r\}_{r\in\mathcal{R}}$,
energy-per-bit evaluation routine $E_{bc}^k(\mathcal{S})$ for any $\mathcal{S}\subseteq\mathcal{R}_0$
\ENSURE Serving sets $\{\mathcal{S}_k\}_{k\in\mathcal{K}}$, infeasible user set $\mathcal{K}_{\mathrm{inf}}$

\FOR{each user $k$}
    \STATE Generate candidate set $\mathcal{C}_k$
    \STATE Compute $E_{bc}^{k\,(i)}$ and $P_k^{(i)}$
    \STATE Discard candidates with $P_k^{(i)}<P_{\min}$
\ENDFOR
\STATE Formulate MILP over $\{y_{k,i}\}$
\STATE Solve using a MILP solver
\STATE Extract relay assignments
\end{algorithmic}
\end{algorithm}

The MILP formulation yields a globally optimal solution over the candidate sets. Any optimality gap arises solely from candidate truncation rather than relaxation, allowing a controlled trade-off between optimality and computational complexity.

Allowing multi-relay cooperation fundamentally changes the structure of the optimization problem by introducing combinatorial coupling among relay selections and nonlinear interactions in the energy-per-bit metric. As a result, the problem becomes NP-hard even for moderate network sizes. To establish performance benchmarks and assess the quality of scalable algorithms, an \emph{exhaustive search} solution is included that evaluates all feasible relay subsets for each user (and all joint assignments when relay capacities are constrained). Although the \emph{exhaustive search} algorithm is computationally infeasible for realistic network dimensions, it provides an exact global optimum for small-scale instances and serves as a reference baseline against which the proposed constructive greedy approximation and optimization-based MILP formulation are evaluated.

\begin{table*}[t]
\centering
\caption{Computational Complexity Summary of Relay Assignment Methods}
\renewcommand{\arraystretch}{1.2}
\begin{tabular}{lll}
\hline
\textbf{Case / Method} & \textbf{Optimality} & \textbf{Complexity} \\ \hline
Case~1(a) Greedy & Optimal & $\mathcal{O}\!\left(K(R+1)\right)$ \\
Case~1(b) Naïve Greedy & Sub-optimal & $\mathcal{O}\!\left(K(R+1)\log(R+1)\right)$ \\
Case~1(b) Min-Cost Flow & Optimal & $\mathcal{O}\!\left(K^2R\log(K+R)\right)$ \\
Case~2 Exhaustive Search & Optimal & $\mathcal{O}\!\left(K2^{R+1}\right)$ \\
Case~2 Greedy Constructive & Sub-optimal & $\mathcal{O}\!\left(KR^2\right)$ \\
Case~2 MILP-Based & Near-optimal & Exponential (candidate-dependent) \\
\hline
\end{tabular}
\label{tab:complexity_summary}
\end{table*}

Table~\ref{tab:complexity_summary} summarizes the optimality–complexity tradeoff across methods. Greedy assignment offers the best scalability, the min-cost flow formulation gives exact polynomial-time solutions for capacitated single-relay assignment, and the MILP formulation provides near-optimal multi-relay solutions at higher cost. Exhaustive search is exact but feasible only for small networks and is used solely as a benchmark. These trends match the empirical results, and the methods suit different operating regimes accordingly. The constructive greedy approximation may incur an optimality gap under tight capacity constraints, whereas MILP controls this gap through its candidate-set size.

\section{Experiments \& Results}
\label{sec:results}

In this section, the proposed relay selection strategies are evaluated under the four settings of Section~\ref{sec:problem-form}, compared in terms of (i) total and average energy per bit and (ii) the number of infeasible users, i.e., those whose received power falls below $P_{\min}$. Results are averaged over $100$ independent random network realizations unless otherwise stated.

Simulations use the mmWave regime with carrier frequency $f_c=29$~GHz and bandwidth $B=100$~MHz, within the licensed mmWave bands targeted for 5G and early 6G. The minimum received power threshold is set to $P_{\min}=-60$~dBm, consistent with practical wideband mmWave receiver sensitivities once directional antenna gains and hardware noise figures are accounted for; this keeps feasibility nontrivial for cell-edge users and highlights the role of relay cooperation in overcoming mmWave path loss. Unless otherwise specified, hardware and link parameters, transmit powers, antenna gains, waste factors, non-path power, and path-loss coefficients, are drawn randomly per realization to model heterogeneous devices. Users are uniformly distributed over a $400\times400~\text{m}^2$ area, with relays placed in a $200\times200~\text{m}^2$ region centered at the base station.

\subsection{Problem Scenarios and Experimental Setup}
\label{sec:scenarios}
\begin{figure*}[t]
    \centering
    \begin{subfigure}[t]{0.35\linewidth}
        \centering
        \includegraphics[width=\linewidth]{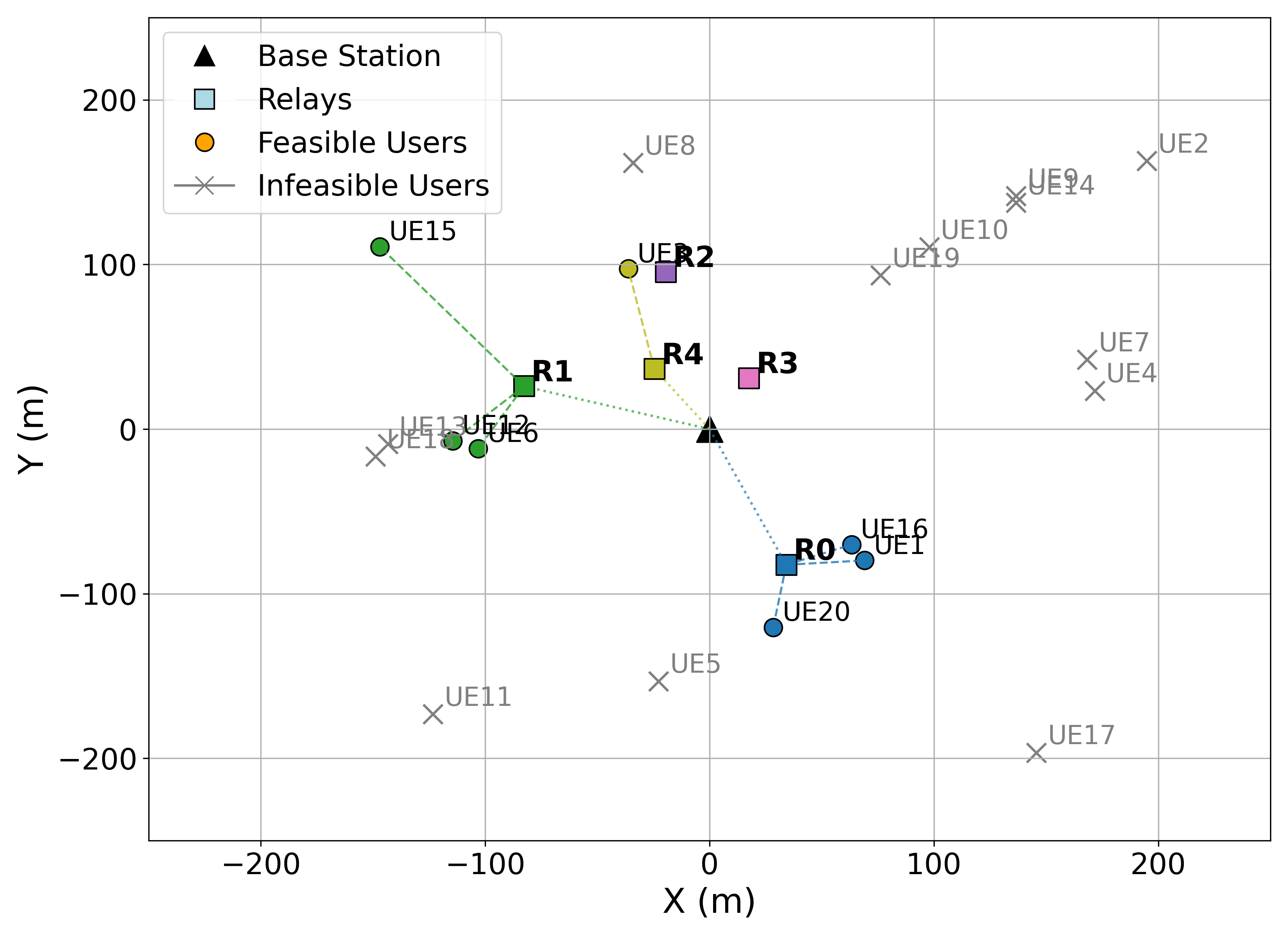}
        \caption{Single-link optimal.}
        \label{fig:motivation_singlelink}
    \end{subfigure}
    \hfill
    \begin{subfigure}[t]{0.35\linewidth}
        \centering
        \includegraphics[width=\linewidth]{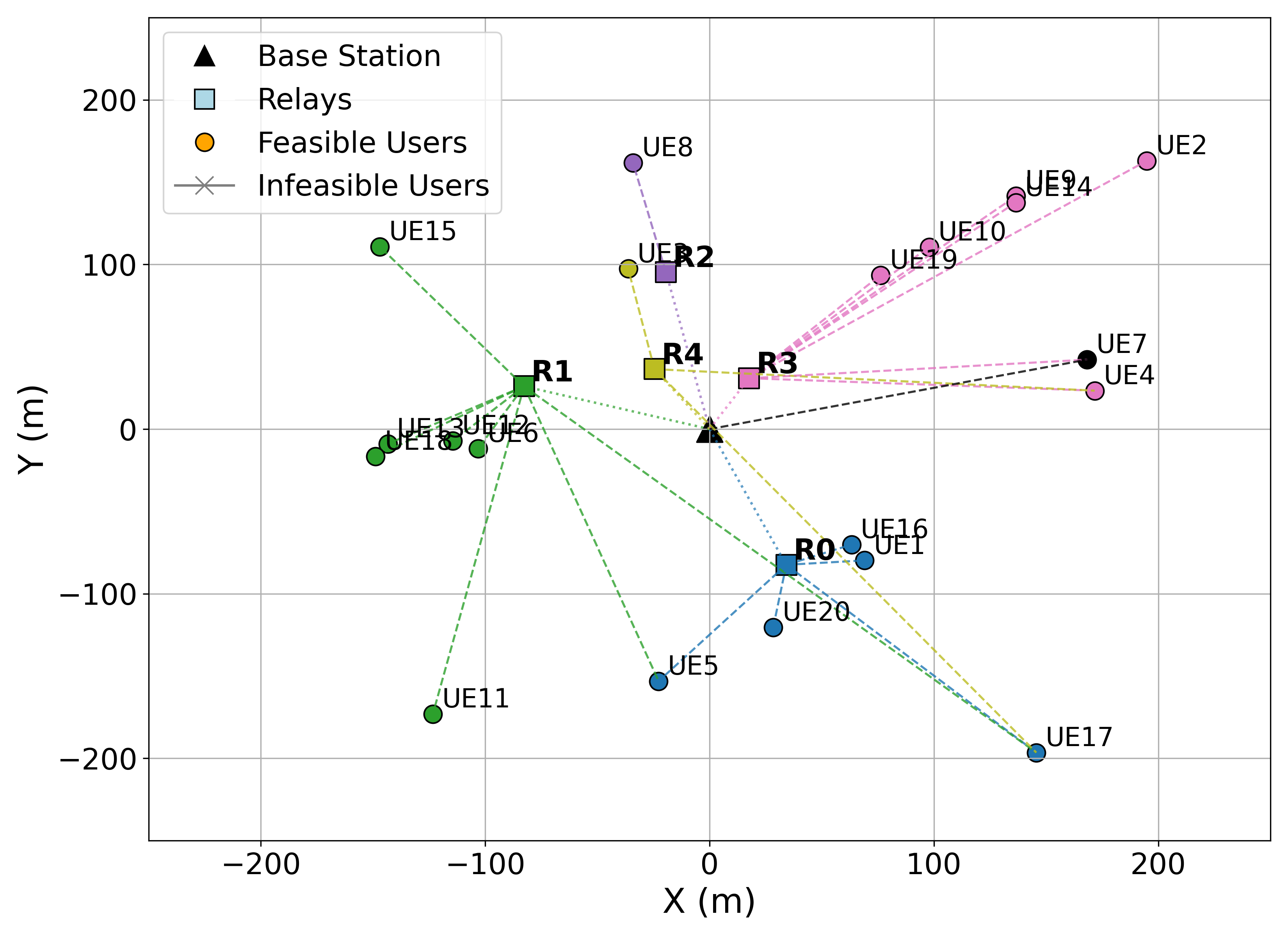}
        \caption{Multi-relay optimal.}
        \label{fig:motivation_multilink}
    \end{subfigure}
    \caption{Motivation for multi-relay cooperation under received power constraints. While single-link serving can fail to satisfy $P_{\min}$, cooperative multi-relay transmission enables power aggregation and significantly reduces the number of infeasible users.}
    \label{fig:motivation_multirelay}
\end{figure*}

The four assignment cases follow the two design dimensions introduced in Section~\ref{sec:problem-form}: single- versus multi-relay serving, and unlimited versus finite relay capacity. Cases~1(a) and~1(b) impose no received-power constraint, since service is always guaranteed by the BS; feasibility therefore becomes relevant only in Cases~2(a) and~2(b), where the minimum received power $P_{\min}$ must be met. The motivation for multi-relay cooperation is illustrated in Fig.~\ref{fig:motivation_multirelay}: under a received-power constraint, single-link serving (Fig.~\ref{fig:motivation_singlelink}) leaves several cell-edge users infeasible because no single relay or direct link delivers $P_{\min}$, whereas multi-relay cooperation (Fig.~\ref{fig:motivation_multilink}) aggregates power across spatially distributed relays and serves most of them. All reported results are averaged over 100 independent random network realizations.

\subsection{Results}

\subsubsection{Case~1(a): Single-Relay Assignment Without Capacity Limits}

\begin{figure}[h]
    \centering
    \includegraphics[width=0.3\linewidth]{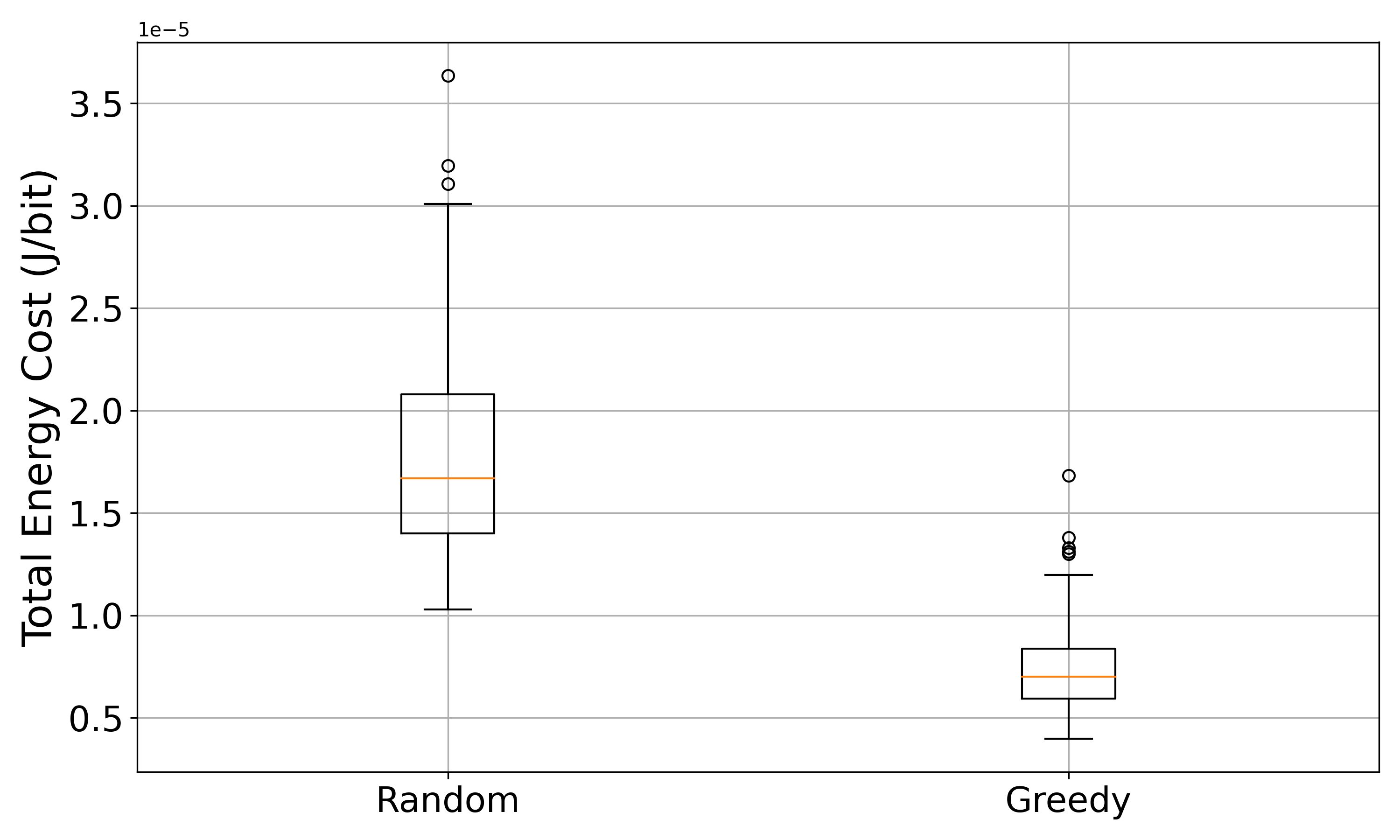}
    \caption{Total energy cost comparison for Case~1(a) over $100$ realizations.}
    \label{fig:case1a}
\end{figure}

Fig.~\ref{fig:case1a} compares the total energy consumption of a random baseline and the proposed greedy selection. The greedy method consistently achieves lower median energy consumption and reduced variability across realizations, confirming that even simple energy-aware selection yields substantial gains when relay capacity is unconstrained.

\subsubsection{Case~1(b): Single-Relay Assignment With Capacity Limits}

\begin{figure}[h]
    \centering
    \includegraphics[width=0.3\linewidth]{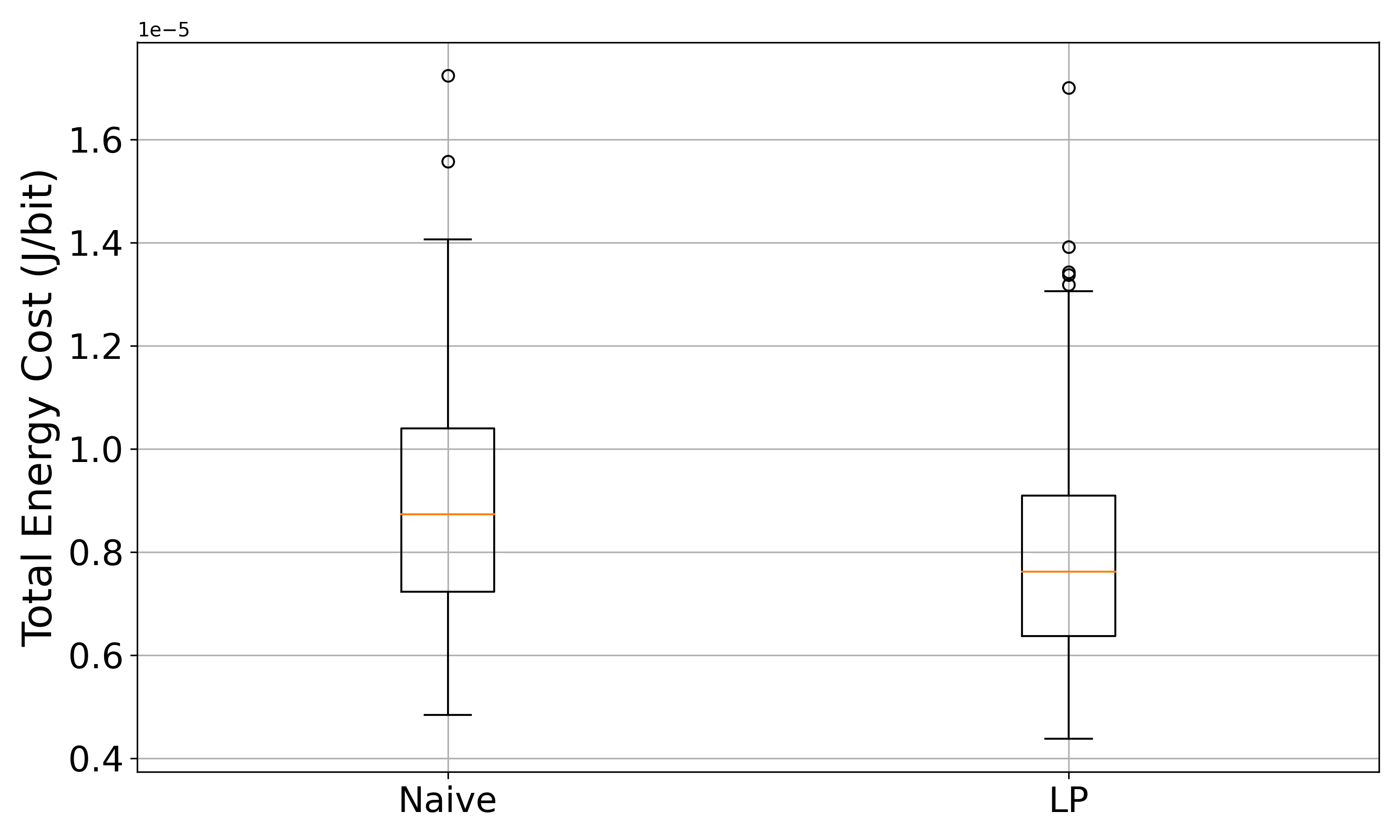}
    \caption{Total energy cost comparison for Case~1(b) with relay capacity limits.}
    \label{fig:case1b_energy}
\end{figure}

Fig.~\ref{fig:case1b_energy} compares the naïve capacity-aware heuristic with the min-cost flow (MCF) formulation. The MCF solution consistently achieves lower energy and reduced variability: unlike the naïve heuristic, which assigns users sequentially without anticipating capacity conflicts, MCF jointly optimizes all assignments under relay capacity constraints, yielding globally optimal solutions in polynomial time.

\subsubsection{Case~2(a): Multi-Relay Assignment Without Capacity Limits}

\begin{figure}[h]
    \centering
    \begin{subfigure}{0.3\linewidth}
        \centering
        \includegraphics[width=\linewidth]{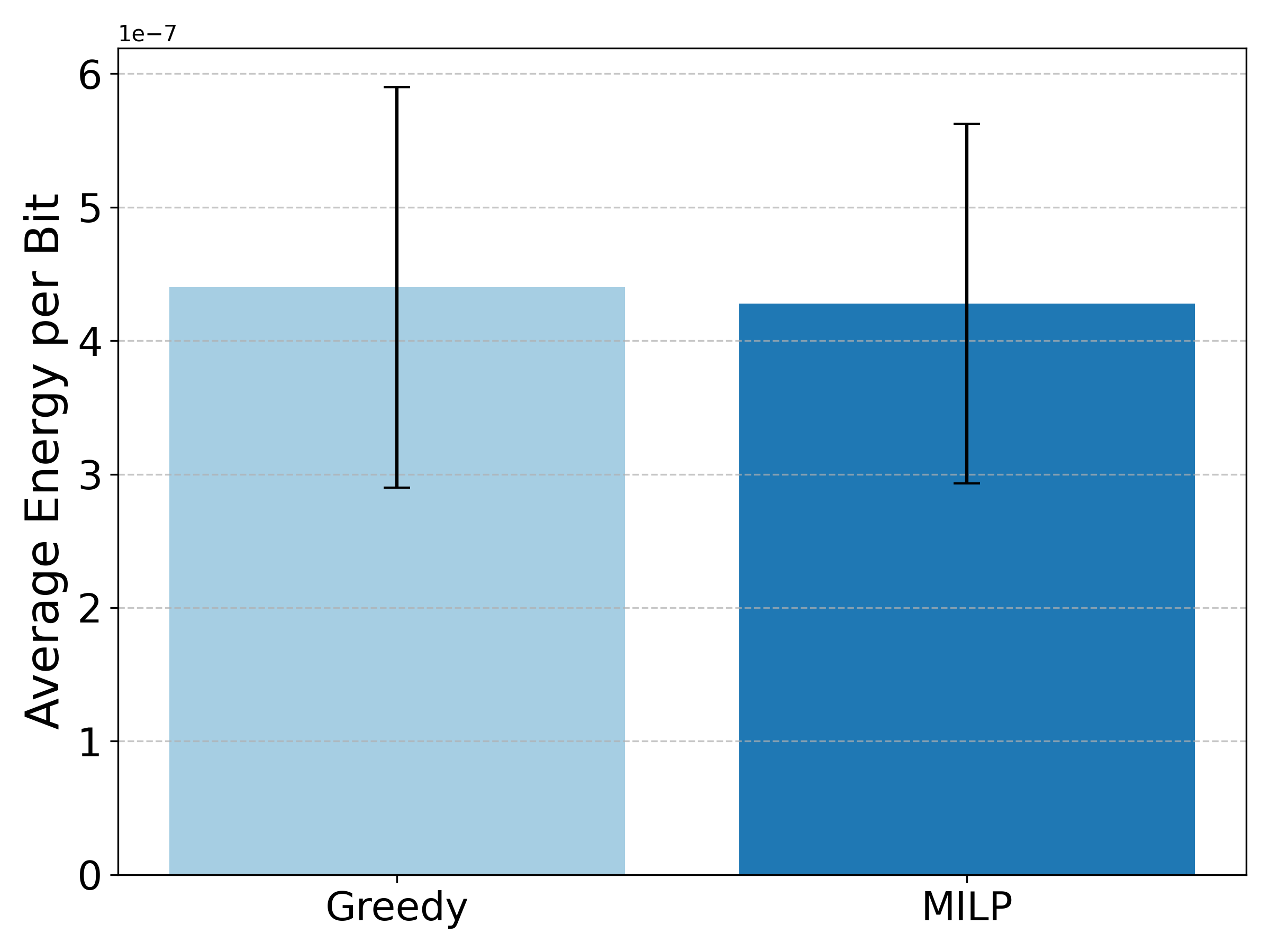}
        \caption{Average energy per bit}
        \label{fig:case2a_energy_large}
    \end{subfigure}
    \hfill
    \begin{subfigure}{0.3\linewidth}
        \centering
        \includegraphics[width=\linewidth]{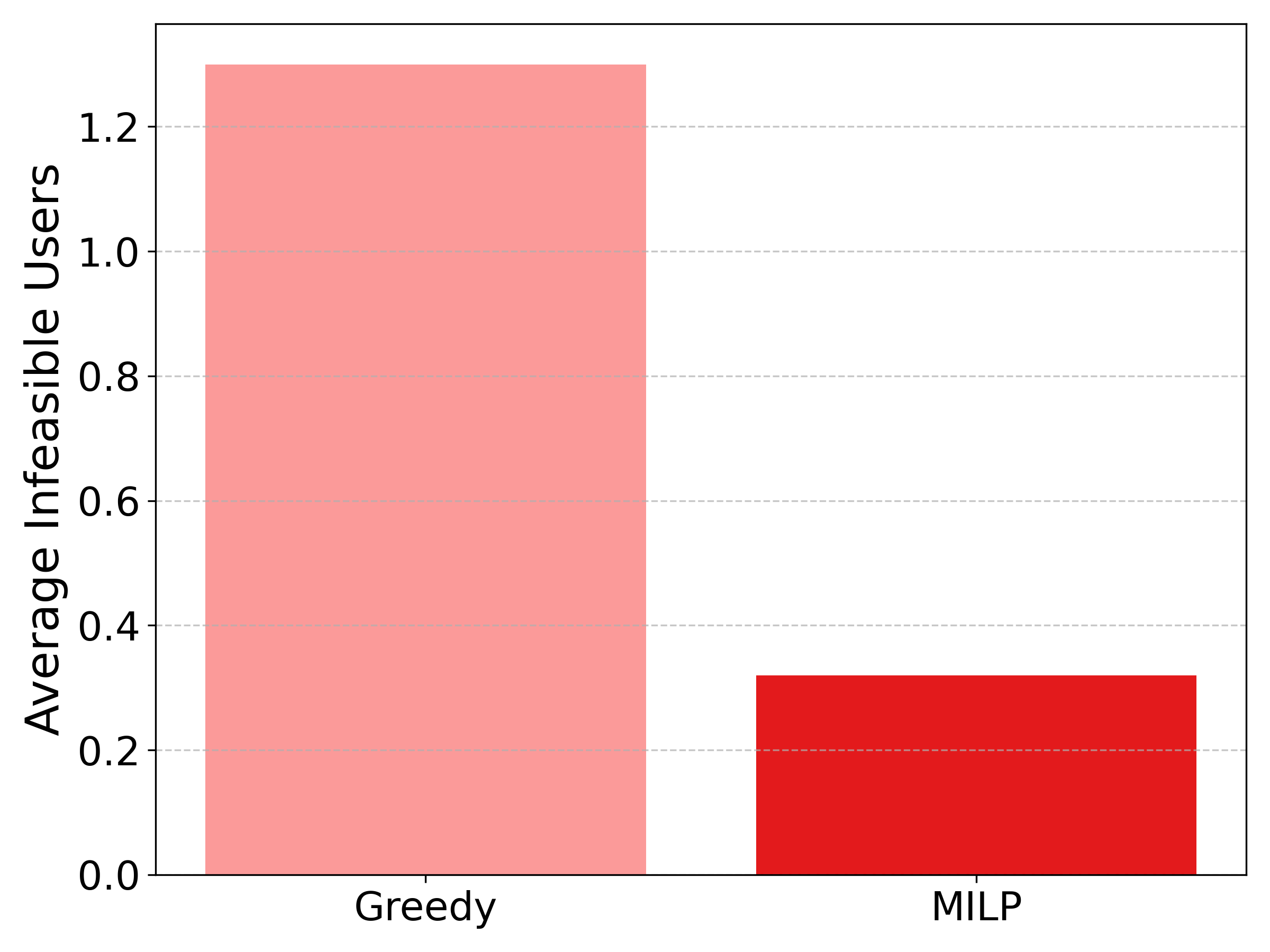}
        \caption{Average infeasible users}
        \label{fig:case2a_infeasible_large}
    \end{subfigure}
    \caption{Case~2(a) performance for $K=20$, $R=5$, $P_{\min}=-60$ dBm.}
    \label{fig:case2a_large}
\end{figure}

\begin{figure}[h]
    \centering
    \begin{subfigure}{0.3\linewidth}
        \centering
        \includegraphics[width=\linewidth]{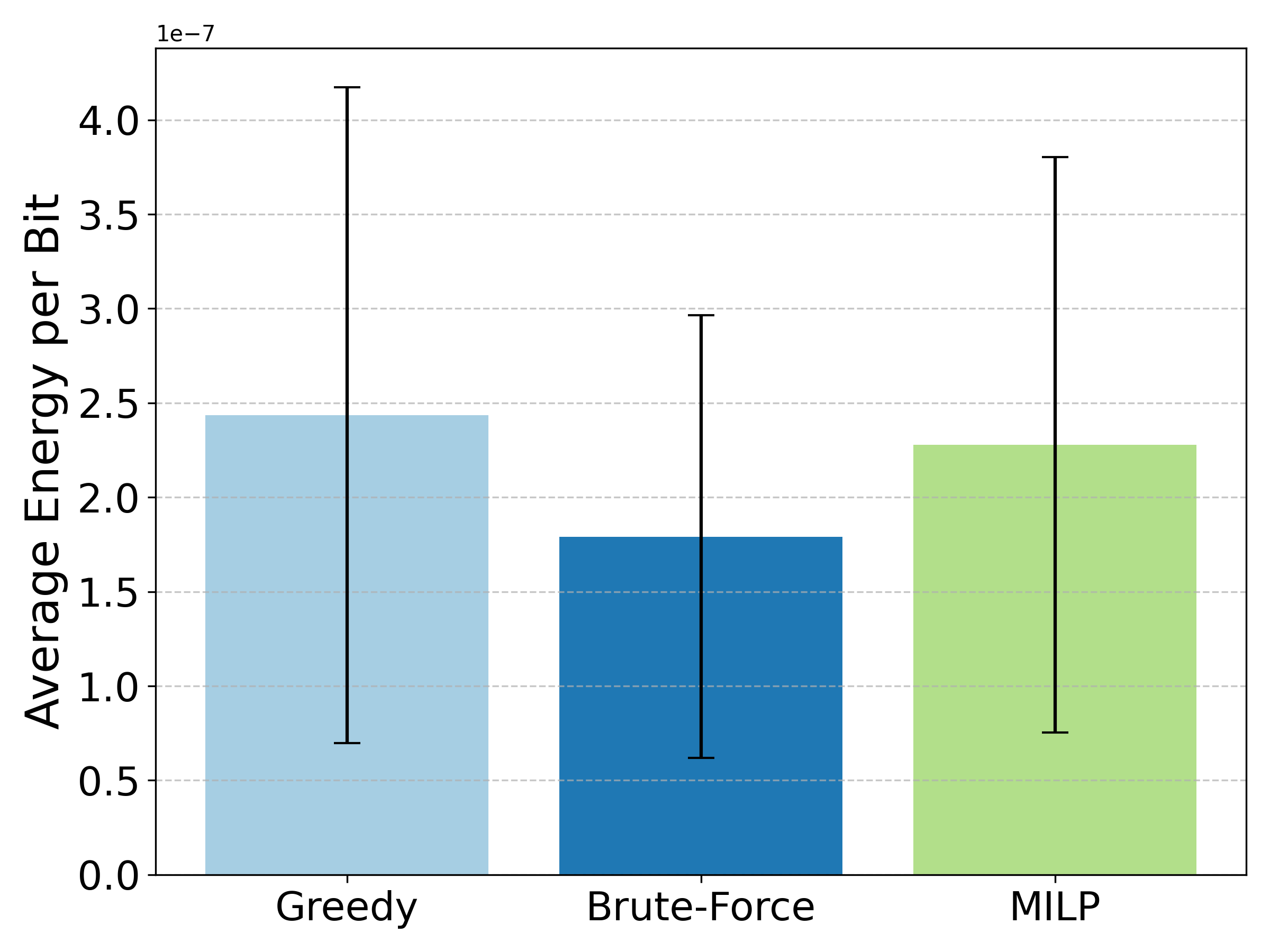}
        \caption{Average energy per bit}
        \label{fig:case2a_energy_small}
    \end{subfigure}
    \hfill
    \begin{subfigure}{0.3\linewidth}
        \centering
        \includegraphics[width=\linewidth]{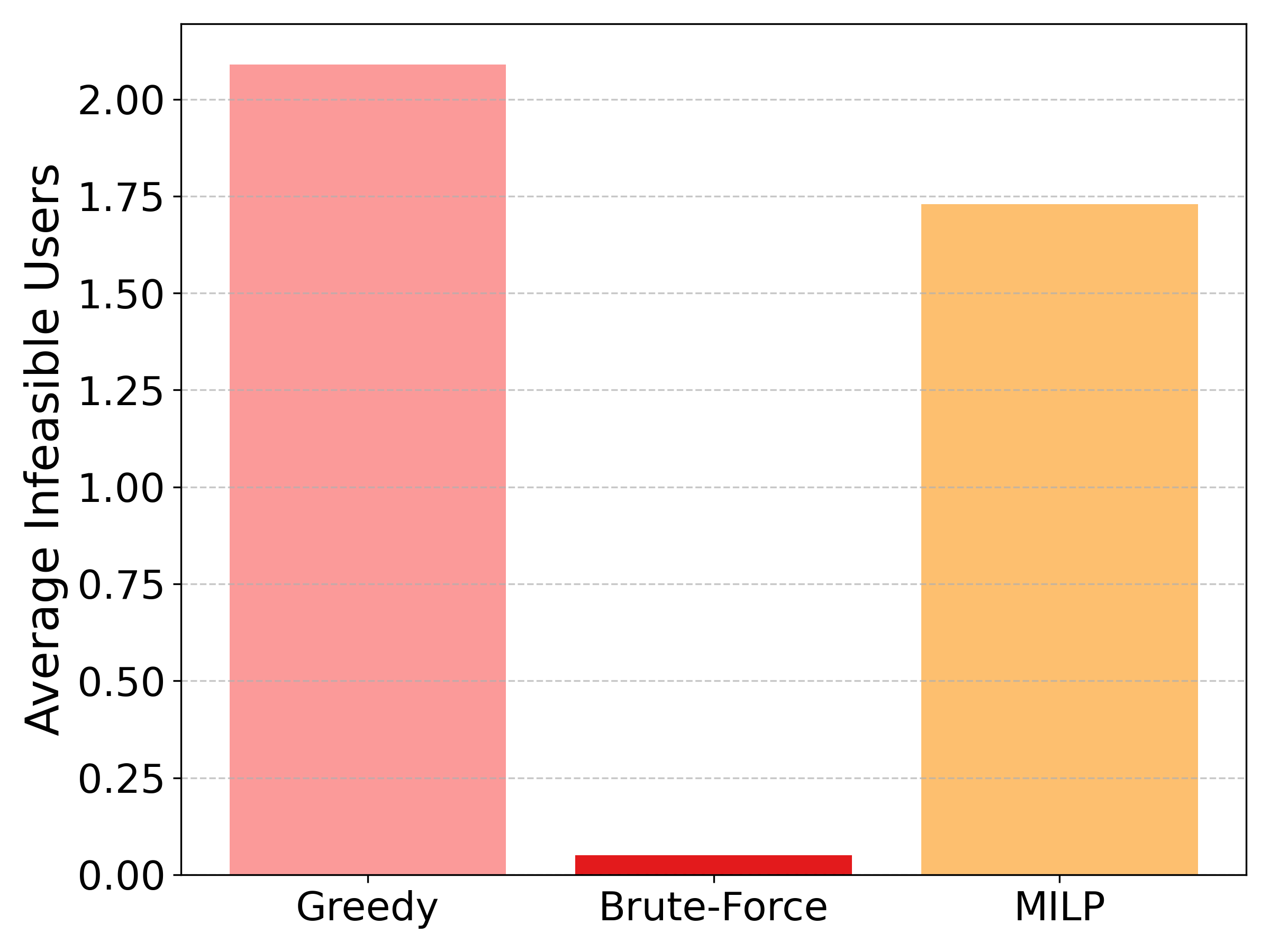}
        \caption{Average infeasible users}
        \label{fig:case2a_infeasible_small}
    \end{subfigure}
    \caption{Case~2(a) performance for $K=5$, $R=3$, $P_{\min}=-60$ dBm.}
    \label{fig:case2a_small}
\end{figure}

Figs.~\ref{fig:case2a_large} and~\ref{fig:case2a_small} report results for large ($K=20$, $R=5$) and small ($K=5$, $R=3$) networks; subfigures~(a) show average energy per bit and~(b) the average number of infeasible users. For the small network, exhaustive search gives the global-optimal reference, with near-zero infeasibility. The greedy and MILP approaches achieve very similar energy costs, confirming that near-optimal performance is attainable at far lower complexity, though greedy leaves slightly more users infeasible. In the large network, exhaustive search is infeasible and omitted. The greedy approach incurs slightly higher energy and substantially more infeasible users than MILP. As no capacity limits apply here, this gap stems from the combinatorial, nonlinear structure of the multi-relay energy model: greedy's incremental local decisions can miss globally efficient combinations, whereas MILP systematically evaluates candidate subsets.

\subsubsection{Case~2(b): Multi-Relay Assignment With Capacity Limits}

\begin{figure}[h]
    \centering
    \begin{subfigure}{0.3\linewidth}
        \centering
        \includegraphics[width=\linewidth]{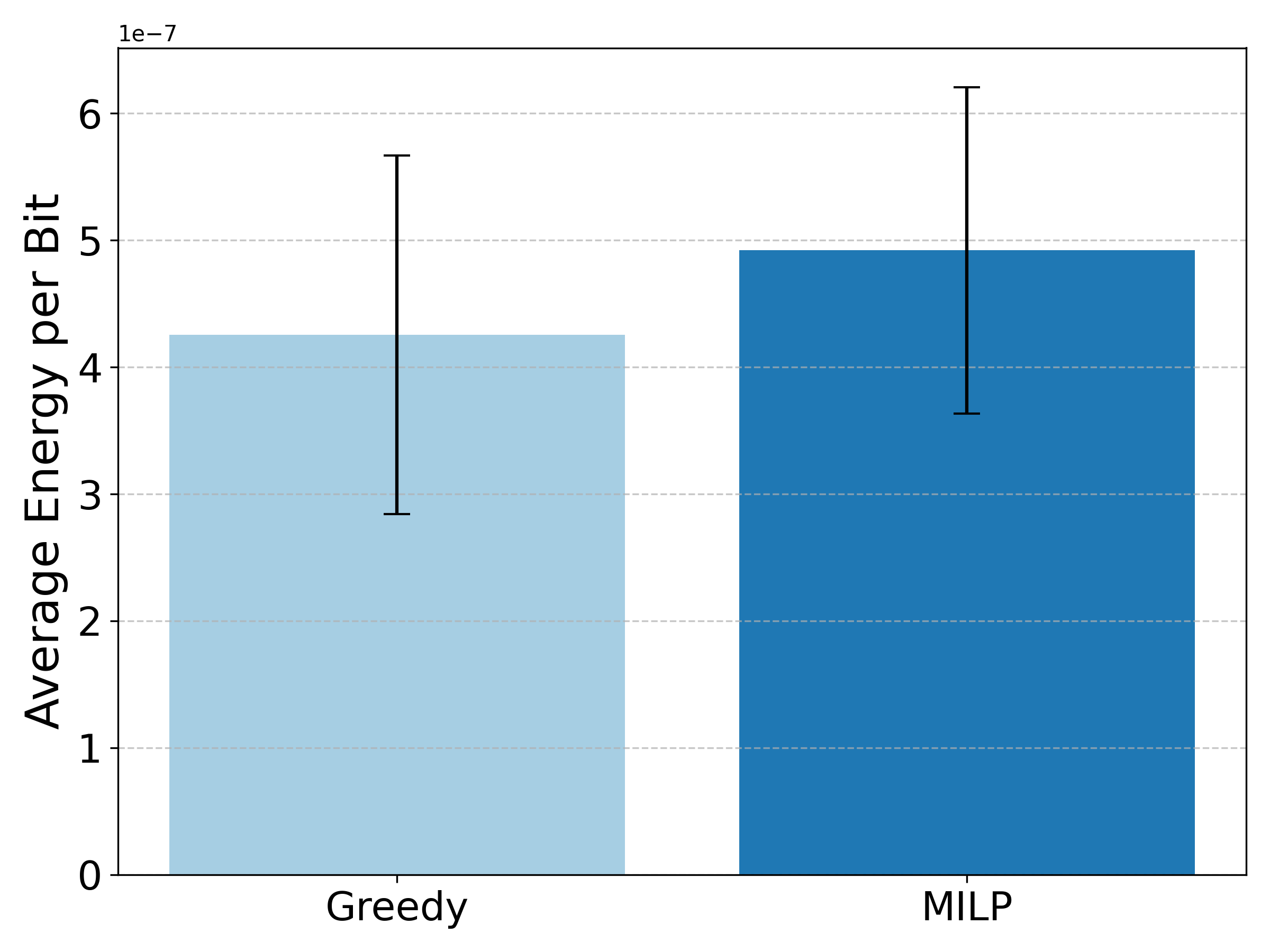}
        \caption{Average energy per bit}
        \label{fig:case2b_energy_large}
    \end{subfigure}
    \hfill
    \begin{subfigure}{0.3\linewidth}
        \centering
        \includegraphics[width=\linewidth]{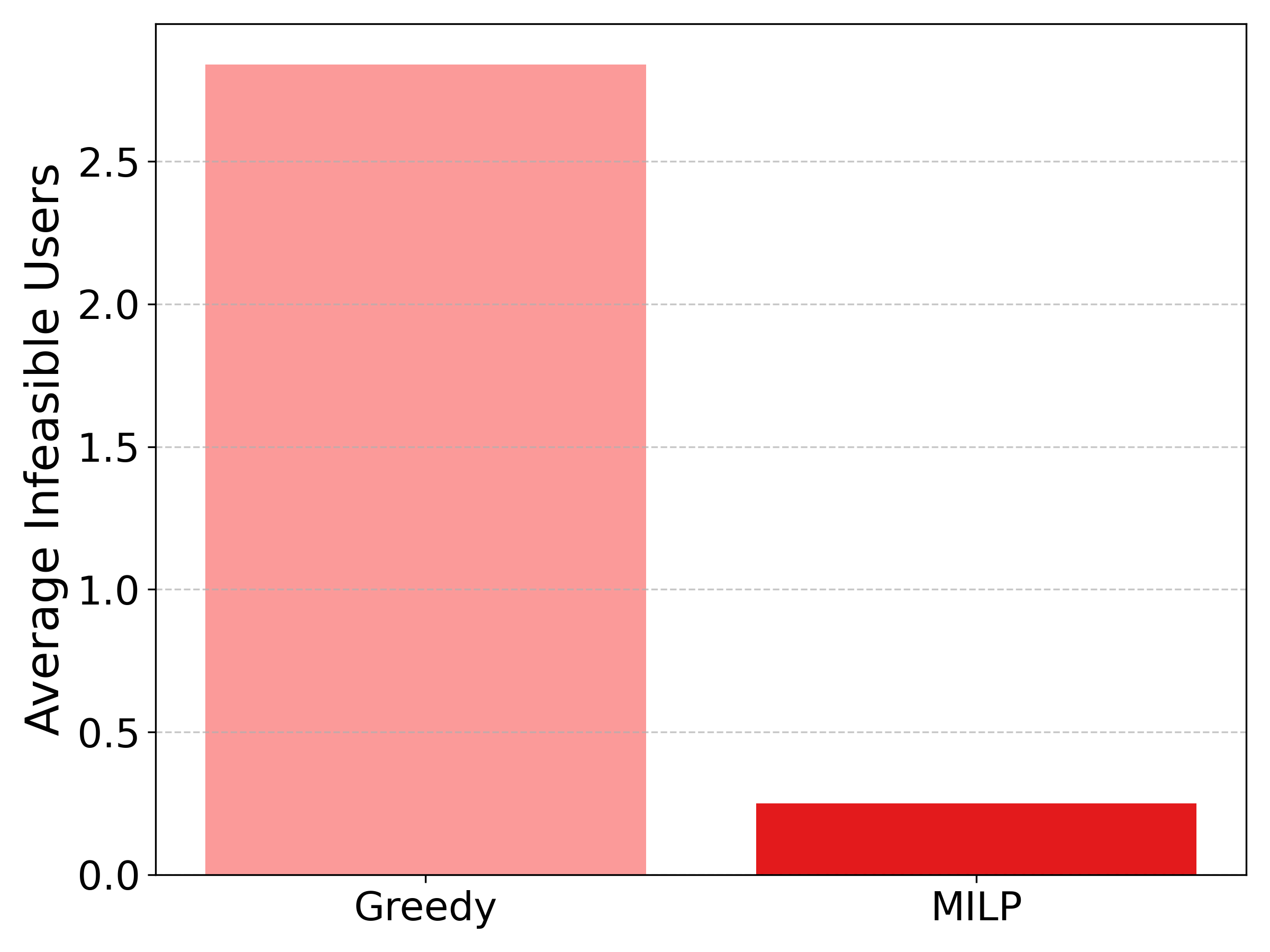}
        \caption{Average infeasible users}
        \label{fig:case2b_infeasible_large}
    \end{subfigure}
    \caption{Case~2(b) performance for $K=20$, $R=5$, $P_{\min}=-60$ dBm.}
    \label{fig:case2b_large}
\end{figure}

\begin{figure}[h]
    \centering
    \begin{subfigure}{0.3\linewidth}
        \centering
        \includegraphics[width=\linewidth]{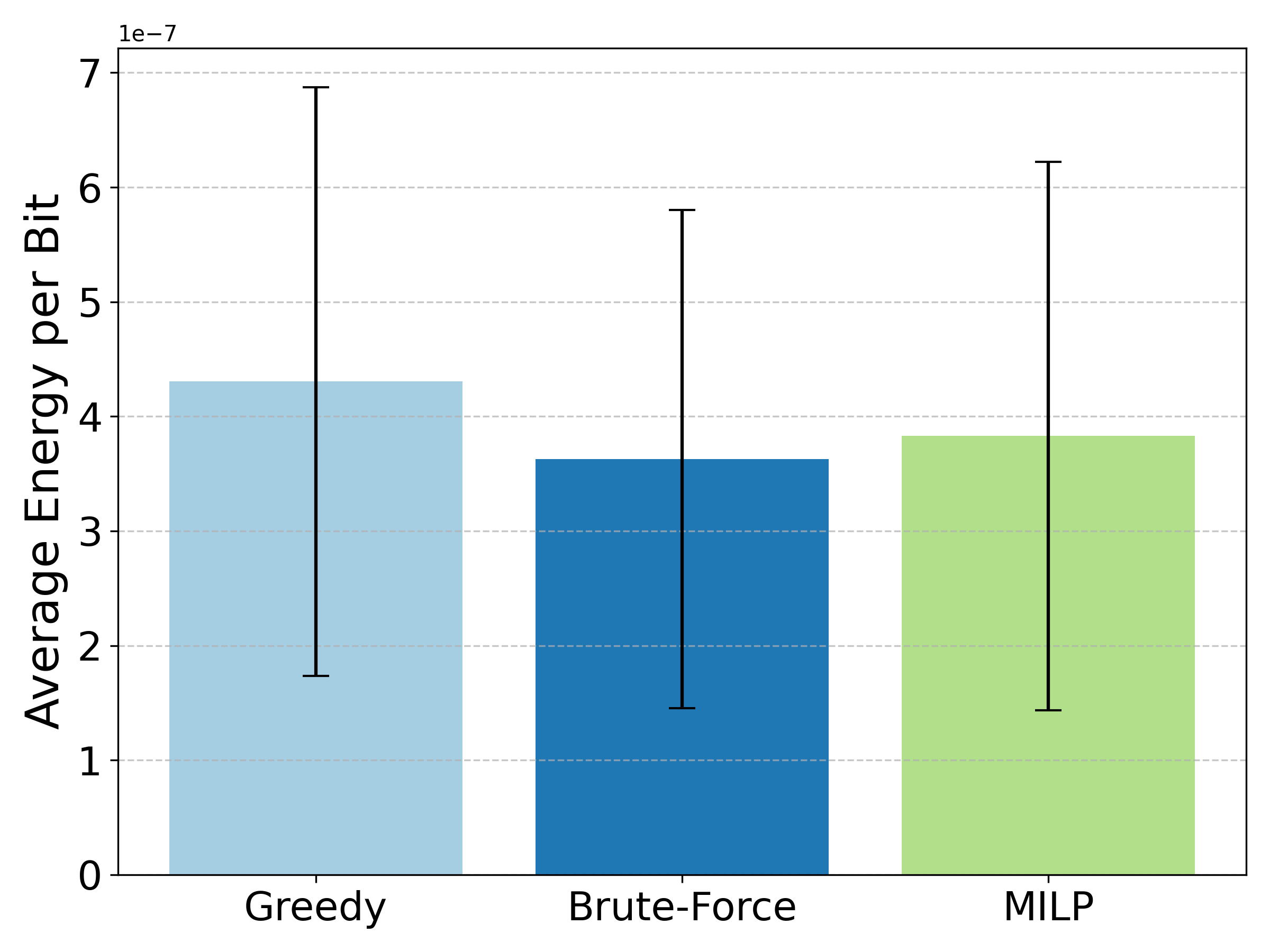}
        \caption{Average energy per bit}
        \label{fig:case2b_energy_small}
    \end{subfigure}
    \hfill
    \begin{subfigure}{0.3\linewidth}
        \centering
        \includegraphics[width=\linewidth]{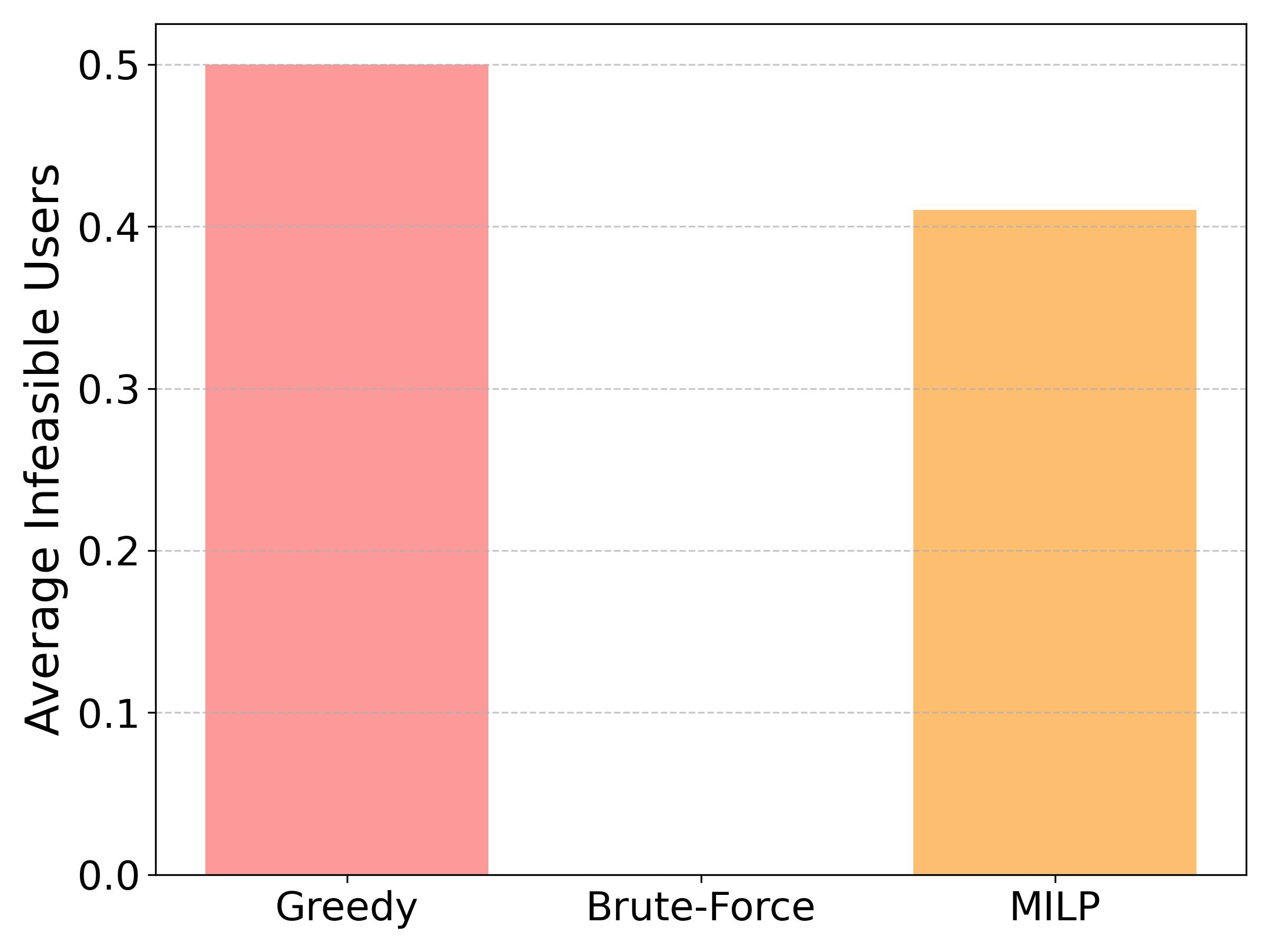}
        \caption{Average infeasible users}
        \label{fig:case2b_infeasible_small}
    \end{subfigure}
    \caption{Case~2(b) performance for $K=5$, $R=3$, $P_{\min}=-60$ dBm.}
    \label{fig:case2b_small}
\end{figure}

Figs.~\ref{fig:case2b_large} and~\ref{fig:case2b_small} summarize the large and small networks; as before, exhaustive search is feasible only at small scale. In the small network, exhaustive search again attains the lowest energy with no infeasible users, while greedy and MILP match in energy but greedy leaves noticeably more users infeasible, reflecting its limited global coordination under capacity constraints. In the large network, greedy's local decisions yield marginally lower energy but more infeasible users, and the capacity coupling makes the gap more pronounced. Case~2(b) is thus the most demanding regime, where achieving both low energy and high feasibility requires coordinated, system-level optimization.

\section{Conclusion}

\label{sec:Conclusion}

This paper developed a Waste Factor (WF)–based framework for hardware-aware energy minimization in multi-user, multi-relay wireless networks. Unlike conventional rate- or transmit-power-based formulations, the proposed framework explicitly accounts for propagation losses, cascaded transceiver inefficiencies, and signal-independent circuit power in a unified end-to-end energy-per-bit metric.

The main result of this work is that WF is not only an energy-efficiency metric, but also a structured modeling framework that makes relay-assignment problems more transparent and computationally tractable. Several structured special cases were shown to admit exact polynomial-time solutions, while the general multi-relay problem remains combinatorial and requires scalable approximations. To address this regime, greedy, min-cost flow, and MILP-based methods were developed to achieve strong energy performance at practical complexity.

The simulation results provide clear evidence for two main system-level outcomes. First, WF-based relay assignment reduces the total energy per bit consumed relative to conventional approaches that rely on proxy objectives, such as rate or transmit power. Second, under minimum received-power constraints, multi-relay cooperation significantly improves coverage feasibility by enabling power aggregation across spatially distributed relays.

Overall, the results show that WF enables both more faithful energy modeling and more effective optimization of relay-assisted wireless systems. This makes the framework particularly valuable for dense and heterogeneous future networks, where hardware inefficiencies and feasibility constraints play a dominant role. Future work will extend the framework to probabilistic propagation effects such as log-normal shadow fading and to site-specific channel modeling.

\ifCLASSOPTIONcaptionsoff
  \newpage
\fi

\bibliographystyle{IEEEtran}
\bibliography{references}

@ARTICLE{rappaport24waste,

  author={Rappaport, Theodore S. and Ying, Mingjun and Piovesan, Nicola and De Domenico, Antonio and Shakya, Dipankar},

  journal={IEEE Open Journal of the Communications Society}, 

  title={Waste Factor and Waste Figure: A Unified Theory for Modeling and Analyzing Wasted Power in Radio Access Networks for Improved Sustainability}, 

  year={2024},

  volume={5},

  number={},

  pages={4839-4867},

  doi={10.1109/OJCOMS.2024.3431459}}

@article{murdock2013consumption,
  title={Consumption factor and power-efficiency factor: A theory for evaluating the energy efficiency of cascaded communication systems},
  author={Murdock, James N and Rappaport, Theodore S},
  journal={IEEE Journal on Selected Areas in Communications},
  volume={32},
  number={2},
  pages={221--236},
  year={2013},
  publisher={IEEE}
}

@article{friis1944noise,
  title={Noise figures of radio receivers},
  author={Friis, Harald T},
  journal={Proceedings of the IRE},
  volume={32},
  number={7},
  pages={419--422},
  year={1944},
  publisher={IEEE}
}

@inproceedings{ying2023waste,
  title={Waste factor: A new metric for evaluating power efficiency in any cascade},
  author={Ying, Mingjun and Shakya, Dipankar and Poddar, Hitesh and Rappaport, Theodore S},
  booktitle={GLOBECOM 2023-2023 IEEE Global Communications Conference},
  pages={6735--6740},
  year={2023},
  organization={IEEE}
}

@ARTICLE{9810097,

  author={Patterson, David and Gonzalez, Joseph and Hölzle, Urs and Le, Quoc and Liang, Chen and Munguia, Lluis-Miquel and Rothchild, Daniel and So, David R. and Texier, Maud and Dean, Jeff},

  journal={Computer}, 

  title={The Carbon Footprint of Machine Learning Training Will Plateau, Then Shrink}, 

  year={2022},

  volume={55},

  number={7},

  pages={18-28},

  doi={10.1109/MC.2022.3148714}}

@online{iea_data_centres,
  author       = {{International Energy Agency}},
  title        = {Data Centres and Data Transmission Networks},
  year         = {2023},
  url          = {https://www.iea.org/energy-system/buildings/data-centres-and-data-transmission-networks},
  note         = {Accessed: 2025-04-13}
}

@techreport{iea2025energyai,
  author       = {{International Energy Agency}},
  title        = {Energy and {AI}},
  institution  = {International Energy Agency},
  year         = {2025},
  url          = {https://www.iea.org/reports/energy-and-ai},
  note         = {Accessed: 2025-04-14}
}

@article{sevim2025theoretical,
  title={Theoretical foundations of waste factor and waste figure with applications to fixed wireless access and relay systems},
  author={Sevim, Nurullah and Ibrahim, Mostafa and Ekin, Sabit and Rappaport, Theodore S},
  journal={npj Wireless Technology},
  volume={1},
  number={1},
  pages={2},
  year={2025},
  publisher={Nature Publishing Group UK London}
}

@inproceedings{ying2024using,
  title={Using waste factor to optimize energy efficiency in multiple-input single-output (MISO) and multiple-input multiple-output (MIMO) systems},
  author={Ying, Mingjun and Shakya, Dipankar and Rappaport, Theodore S},
  booktitle={GLOBECOM 2024-2024 IEEE Global Communications Conference},
  pages={3170--3175},
  year={2024},
  organization={IEEE}
}

@ARTICLE{jiang19uav,
  author={Jiang, Xu and Wu, Zhilu and Yin, Zhendong and Yang, Zhutian},
  journal={IEEE Wireless Communications Letters}, 
  title={Joint Power and Trajectory Design for UAV-Relayed Wireless Systems}, 
  year={2019},
  volume={8},
  number={3},
  pages={697-700},
  doi={10.1109/LWC.2018.2885056}}

@ARTICLE{cai08afrelay,
  author={Cai, Jun and Shen, Xuemin and Mark, Jon W. and Alfa, Attahiru S.},
  journal={IEEE Transactions on Wireless Communications}, 
  title={Semi-Distributed User Relaying Algorithm for Amplify-and-Forward Wireless Relay Networks}, 
  year={2008},
  volume={7},
  number={4},
  pages={1348-1357},
  doi={10.1109/TWC.2008.060909}}

@INPROCEEDINGS{thai11coordianted,
  author={Thai, Chan Dai Truyen and Popovski, Petar and Kaneko, Megumi and de Carvalho, Elisabeth},
  booktitle={IEEE International Conference on Communications (ICC)}, 
  title={Coordinated Transmissions to Direct and Relayed Users in Wireless Cellular Systems}, 
  year={2011},
  volume={},
  number={},
  pages={1-5},
  doi={10.1109/icc.2011.5962696}}

@ARTICLE{irmer2011comp,
  author={Irmer, Ralf and Droste, Heinz and Marsch, Patrick and Grieger, Michael and Fettweis, Gerhard and Brueck, Stefan and Mayer, Hans-Peter and Thiele, Lars and Jungnickel, Volker},
  journal={IEEE Communications Magazine}, 
  title={Coordinated multipoint: Concepts, performance, and field trial results}, 
  year={2011},
  volume={49},
  number={2},
  pages={102-111},
  doi={10.1109/MCOM.2011.5706317}}

@ARTICLE{gesbert2010comp,
  author={Gesbert, David and Hanly, Stephen and Huang, Howard and Shamai Shitz, Shlomo and Simeone, Osvaldo and Yu, Wei},
  journal={IEEE Journal on Selected Areas in Communications}, 
  title={Multi-Cell MIMO Cooperative Networks: A New Look at Interference}, 
  year={2010},
  volume={28},
  number={9},
  pages={1380-1408},
  doi={10.1109/JSAC.2010.101202}}

@ARTICLE{zhao2022comp,
  author={Zhao, Junhui and Yang, Lihua and Xia, Minghua and Motani, Mehul},
  journal={IEEE Internet of Things Journal}, 
  title={Unified Analysis of Coordinated Multipoint Transmissions in mmWave Cellular Networks}, 
  year={2022},
  volume={9},
  number={14},
  pages={12166-12180},
  doi={10.1109/JIOT.2021.3134017}}

@ARTICLE{li2019comp,
  author={Li, Zhao and Chen, Jie and Zhen, Lu and Cui, Sha and Shin, Kang G. and Liu, Jia},
  journal={IEEE Transactions on Wireless Communications}, 
  title={Coordinated Multi-Point Transmissions Based on Interference Alignment and Neutralization}, 
  year={2019},
  volume={18},
  number={7},
  pages={3347-3365},
  doi={10.1109/TWC.2019.2908159}}

@ARTICLE{cover79relay,
  author={Cover, T. and Gamal, A.E.},
  journal={IEEE Transactions on Information Theory}, 
  title={Capacity theorems for the relay channel}, 
  year={1979},
  volume={25},
  number={5},
  pages={572-584},
  doi={10.1109/TIT.1979.1056084}}

@article{rappaport2024wastemj,
  title={Waste figure and waste factor: New metrics for evaluating power efficiency in any circuit or cascade},
  author={Rappaport, Theodore and Ying, Mingjun and Shakya, Dipankar},
  journal={Microwave Journal},
  volume={67},
  number={5},
  pages={54--84},
  year={2024}
}

\end{document}